 \documentclass[preprint]{aastex}

\shorttitle{Speckle Measures with DSSI. V.}
\shortauthors{Horch et al.}

\begin{document}
\newcommand{\beq}{\begin{equation}}
\newcommand{\eeq}{\end{equation}}
\newcommand{\ea}{{\it et al.\ }}


\title{Observations of Binary Stars with the Differential Speckle
Survey Instrument. V. 
Toward an Empirical Metal-Poor Mass-Luminosity Relation}

\author{Elliott P. Horch\altaffilmark{1,8,9},
William van Altena\altaffilmark{2}, 
Pierre Demarque\altaffilmark{2},
Steve B. Howell\altaffilmark{3,8},
Mark E. Everett\altaffilmark{4,8},
David R. Ciardi\altaffilmark{5,8},
Johanna K. Teske\altaffilmark{6,8},
Todd J. Henry\altaffilmark{7},
and Jennifer G. Winters\altaffilmark{7}}

\affil{$^{1}$Department of Physics, 
Southern Connecticut State University,
501 Crescent Street, New Haven, CT 06515}

\affil{$^{2}$Department of Astronomy, Yale University
P.O. Box 208101, New Haven, CT 06520-8101}

\affil{$^{3}$NASA Ames Research Center, Moffett Field, CA 94035}

\affil{$^{4}$National Optical Astronomy Observatory,
950 N. Cherry Ave, Tucson, AZ 85719}

\affil{$^{5}$NASA Exoplanet Science Institute, California Institute 
of Technology, 770 South Wilson Avenue, Mail Code 100-22, Pasadena, CA 91125}

\affil{$^{6}$Department of Terrestrial Magnetism and Carnegie Observatories, 
Carnegie Institute of 
Washington, 5241 Broad Branch Road, NW,
Washington, DC 20015} 

\affil{$^{7}$Department of Physics and Astronomy, Georgia State University,
Atlanta, GA 30302}

\email{horche2@southernct.edu,
william.vanaltena@yale.edu, pierre.demarque@yale.edu,
steve.b.howell@nasa.gov,
everett@noao.edu,
ciardi@ipac.caltech.edu,
jteske@carnegiescience.edu,
thenry@astro.gsu.edu,
winters@astro.gsu.edu}

\altaffiltext{8}{Visiting Astronomer, Gemini Observatory,
National Optical Astronomy Observatory,
which is operated by the Association of
Universities for Research in Astronomy, Inc., under a cooperative agreement 
with the NSF on behalf of
the Gemini partnership: the National Science Foundation (United States), 
the Science and Technology
Facilities Council (United Kingdom), 
the National Research Council (Canada), CONICYT (Chile), the
Australian Research Council (Australia), Minist\'{e}rio da Ci\'{e}ncia, 
Tecnologia e Inova\~{n}ao (Brazil) and Ministerio
de Ciencia, Tecnolog\'{\i}a e Innovaci\'{o}n Productiva (Argentina)}

\altaffiltext{9}{Adjunct Astronomer, Lowell Observatory.}


\begin{abstract}

In an effort to better understand the details of the stellar structure and evolution
of metal poor stars, the Gemini North telescope was used on two occasions to take speckle
imaging data of a sample of known spectroscopic binary stars and 
other nearby stars in order to search for and resolve close companions.
The observations were obtained using the Differential Speckle Survey 
Instrument, which takes data in two filters simultaneously.
The results presented here are of 90 observations of 23 systems in which
one or more companions was detected, and 6 stars where no companion was 
detected to the limit of the camera capabilities at Gemini. In the case of
the binary and multiple stars, these results are then further analyzed to
make first orbit determinations in five cases, and orbit refinements
in four other cases. Mass information is derived, and since the systems
span a range in metallicity, a study is presented that compares 
our results with the expected
trend in total mass as derived from the most 
recent Yale isochrones as a function
of metal abundance. These data
suggest that metal-poor main-sequence stars are less massive at a given color
than their solar-metallicity analogues in a manner consistent with that
predicted from the theory.

\end{abstract}


\section{Introduction}

Torres \ea (2010) have published the most definitive information we 
have on the mass-luminosity relation (MLR) for main sequence stars to 
date, but the sample of systems that has yielded the best masses contains 
only one system with a measured metal abundance ([m/H]) 
of less than -0.25. More recent 
studies using long baseline optical interferometry data have begun to 
address metallicity ({\it e.g.\ }Boyajian 2012a, 2012b, 
Freiden and Chaboyer 2012), but only as low as about [m/H]=-0.5. 
Masses and luminosities of metal-poor stars are extremely important to 
calibrate precisely. For example, the Population II main sequence has been 
defined by nearby metal-poor stars ({\it e.g.\ }Reid 1997, Gratton \ea 1997), a 
number of which may be binary. If metal-poor binaries are resolved and individual 
luminosities can be obtained, these new data could be used to reduce 
the current scatter in the metal-poor main sequence, 
allowing for more stringent constraints on stellar 
models, as well as better ages and distances to galactic globular 
clusters. The secondary components of metal-poor binaries are 
especially important in that they will have undergone considerably 
less change in color and luminosity, and their current observables 
should thus be close to their zero age locations in the color-magnitude 
diagram. 

Unfortunately,
metal-poor systems in the Solar neighborhood are less numerous and also 
typically farther away and therefore fainter than classic Population I systems, 
often making them difficult objects for optical interferometry. In addition, 
determining high-quality individual masses is usually 
time-consuming and requires 
both astrometric and spectroscopic observations that span the orbital 
period. The DSSI speckle camera combined with the Gemini North telescope
provides an excellent opportunity 
to make quick progress on several low-metallicity systems by combining 
definitive high-resolution observations obtained at Gemini with spectroscopic data and 
lower-precision astrometric data taken at other telescopes that is
already in the literature.

Direct empirical determinations of individual masses of spectroscopic 
binaries are possible if the components can be resolved and the separation 
between the components can be accurately determined. For double-lined 
systems, the combined spectroscopic/astrometric orbit solution yields 
individual masses as well as a distance to the system without recourse 
to parallax measurements; if the distance is independently known, then this can 
in principle be 
used to further constrain the mass information. For single-lined systems, 
an independent distance measure is needed to complete the path to 
individual masses. With a spectroscopic orbit and parallax in hand, 
even two or three resolved observations spread out along the orbit can 
be used to reliably measure the semi-major axis and inclination 
and therefore provide the basis for mass determinations.

The observations presented in this paper include stars that
span a range of metal abundance from approximately the solar value 
to [m/H]=-1.39 with roughly one third having
[m/H] values in the range from -0.4 to -1.39. 
The spectral types for the sample as a whole range from early-F to early-K. 
We selected the list of targets from two main sources: the sample of 
double-lined spectroscopic binaries appearing in the Geneva-Copenhagen
spectroscopic survey of nearby stars (Nordstr\"{o}m \ea 2004), and the 
spectroscopic survey of proper motion stars of Carney, Latham, 
and their collaborators (e.g.\ Goldberg \ea 2002 and references
therein). We looked for stars that also had revised {\it Hipparcos}
parallaxes (van Leeuven 2007)
and roughly prioritized the resulting subset by a combination
of distance
and metal abundance. 
Most of the systems that we observed already have spectroscopic
orbits. With the astrometry obtained in the work described here,
we can derive
mass information and constrain stellar structure theory; in 
particular, the dependence of mass on metallicity
for a given spectral type (or equivalently, effective temperature) on the main sequence.
While our ultimate goal is to provide a high-precision empirical
calibration of the low-metallicity mass-luminosity relation 
and to use the most relevant systems to obtain 
a detailed understanding of metal-poor
stellar evolution, 
the current work centers on the identification of some of those systems
most likely to provide important information and a preliminary study
of the trend in mass with metallicity.

\section{Observations and Data Reduction}

The Differential Speckle Survey Instrument (DSSI, Horch \ea 2009) 
was first used at the Gemini North telescope in July of 2012. Results
of those observations are found in Horch \ea (2012), Howell \ea
(2012), and Horch \ea (2014). In July of 2013 and 2014, DSSI
enjoyed official visiting
instrument status at the Gemini North, where it was made available to
the community, and several scientific 
programs were executed in each of those
two summers.

The instrument records speckle patterns in two filters simultaneously.
We refer to the two channels of the instrument as the reflective and
transmissive channels, depending on whether the light detected has 
been transmitted through the dichroic element or if it has been reflected
off of it. In the case of all observations here, the reflective channel 
recorded data through a filter with a center wavelength of 880 nm
and a 50-nm full width at half maximum (FWHM) transmission, 
and the transmissive channel recorded data through a 692-nm center-wavelength
filter with a 40-nm FWHM. These filters were chosen to maximize the
limiting magnitude and overall data quality obtained with
the instrument. Given that the effects of 
atmospheric turbulence are less severe at redder wavelengths, these
filters allow us to use longer frame integration times on each speckle
pattern. Atmospheric dispersion is less of an issue in the red,
so that the use of these filters permits high-quality observing at larger
airmass. The two wavelengths are also sufficiently separated to give
color information of the components of the binary systems that we observe.

During the 2013 and 2014 runs at Gemini, we were able to obtain observations
of a number of binaries drawn from the lists discussed above, 
as well as to search for close companions to several nearby stars
in other cases. This paper is mainly focused on the results of the former
group, although it was sensible to include the latter one as the data reduction
and analysis techniques were identical. 
In both cases, Gemini's large aperture allows DSSI to obtain
extremely high-resolution images owing to the smaller diffraction limit
of the telescope relative to most other speckle programs in operation
today, as well as the ability to 
successfully observe relatively faint sources
that in many cases are not easily observable at this time using long-baseline
optical interferometry. 

\subsection{Reduction Method}

The reduction scheme for binary star observations with DSSI has been 
described in other papers, most recently in Horch \ea (2011a) and Horch \ea 
(2012).
For the observations discussed here, the typical observation consisted of
a sequence of 1000 60-ms exposures recorded in each channel of the instrument
simultaneously. These are stored as separate FITS stacks, where each frame
has a format of 256$\times$256 pixels. The reduction consists of (1) forming 
the autocorrelation of each frame and summing these over the 1000-frame stack,
and (2) computing the so-called ``near-axis'' subplanes of the image bispectrum
for each observation.
A reconstructed image is then formed by dividing the 
Fourier transform of the autocorrelation of the binary with that of the 
point source, taking the square root, and then combining that with a
phase function derived from the bispectral subplanes using the method
of Meng \ea (1990). (The point source data are obtained by observing a 
bright, unresolved star.) That results in 
an estimate of the Fourier transform of the true, diffraction-limited
source intensity distribution. It is low-pass Gaussian filtered and 
inverse transformed to arrive at the reconstructed image. The reconstructed
image of an observation is the primary data product that we use 
for determining if a companion is present. 
If no companions are seen, then we use the reconstruction to derive 
the detection limits for the observation. If
at least one companion is detected, then in order to obtain the astrometry
and photometry relative to the primary star, we use the power spectrum, where
we perform a weighted least-squares fit to a cosine squared function, 
that is, the fringe pattern seen in the Fourier plane.

For the 2014 run, we developed a program that would allow us greater 
flexibility in the choice of a point source and greater efficiency
while observing. Point sources have generally been necessary to observe 
close in time and close in sky position to our science targets 
in order to have a ``real time'' point spread function that matches
the observing conditions of the science target for our
deconvolution process. (This is especially true at airmasses above 1.4.)
Recognizing that the details of the point spread function are mainly 
due to 
residual atmospheric dispersion and therefore related to sky position
of the source at the time of the observation (altitude and azimuth), the
new program takes as input a point source observed at very high elevation 
({\it i.e.\ }one with little dispersion) and
builds in the expected dispersion for the sky position of the science
target. We compared the results from both point sources taken near in time
and near in sky position to the science target versus those from point
sources made in this way, and found no significant difference in the 
quality of the astrometry and photometry. Generally speaking, for the
objects shown in Table 1, we used unmodified point sources for most of
the 2013 observations and high-elevation point sources 
modified by the program for the 2014 observations.

\subsection{Pixel Scale and Orientation}

The pixel scale and orientation were determined using the same method
that was used in our first experience with DSSI at Gemini (Horch \ea 2012).
While our preferred 
method would have been to use a slit mask mounted in the converging
beam of the telescope as we have done at WIYN,
the practicalities of mounting and unmounting such a mask at Gemini
as well as a desire to make the science observing as efficient as
possible have led us to the use of calibration binaries to derive
the pixel scale. For the
present work,
we selected three bright binaries with extremely high-quality 
orbits appearing in the {\it Sixth Catalog of Visual Orbits of
Binary Stars} (Hartkopf \ea 2001a). These were HIP 83838 = HU 1176AB,
HIP 104858 = STT 535, and HIP 104887 = AGC 13AB. We observed each with
the instrument, reduced the data in the manner described in the previous
subsection, and compared the location of the secondary in the resulting
data with the ephemeris positions in each case.

From our long observing program with DSSI at the WIYN telescope\footnotemark 
\hspace{0pt} with DSSI (2008-2013),
we know that there is a small amount of distortion in the reflective channel;
we have been able to map this out extensively at WIYN, and it has remained 
essentially 
constant throughout the years of use at that telescope. The position angles 
of the binaries used in the
scale calibration for the current work allowed us to determine that
the effect was consistent with the WIYN distortion model; we therefore 
assumed that model and then calculated the final position angles and
separation for the calibrators based on that. We then obtained scale values
of 0.0108 arc seconds per pixel in the transmissive channel of the instrument
(692 nm) and 0.0114 arc seconds per pixel in the reflective channel (880 nm).
Using the published uncertainties in the orbital elements and our own
measurement uncertainties as discussed below, we estimate that these values
are uncertain at the level of approximately $\pm$0.1\%. 
Likewise, the chip orientation is determined to within about $\pm$0.2 degrees.
Given that the speckle images
had a format of 256$\times$256 pixels, the field of view was therefore about
2.8$\times$2.8 arc seconds.

\footnotetext{The WIYN Observatory is a joint facility of
the University of Wisconsin-Madison, Indiana University, 
the National Optical Astronomy Observatory and the University of Missouri.}

\section{Results}

Table 1 shows the main results of the observations. The columns give:
(1) the Washington Double Star (WDS) number (Mason \ea 2001a), which also
gives the right ascension and declination for the object in 2000.0 coordinates;
(2) a secondary identifier, most often
the Henry Draper Catalogue (HD) number for the object;
(3) the Discoverer Designation;
(4) the {\it Hipparcos} Catalogue number (ESA 1997);
(5) the Besselian date of the observation;
(6) the position angle ($\theta$)
of the
secondary star relative to the primary, with North through East defining the
positive sense of $\theta$; (7) the separation of the two stars ($\rho$), in
arc seconds; (8) the magnitude difference ($\Delta m$)
of the pair; (9) center wavelength of the filter used; and
(10) full width at half maximum of the filter transmission in nanometers.
Position angles have not been precessed from the dates shown and are
left as determined by our analysis procedure, even if inconsistent with
previous measures in the literature. Two objects have no previous detection
of the companion; we suggest discoverer designations of DSG (DSSI-Gemini) 
7 and 8, and will refer to them as such throughout the rest of this
paper. DSG 7 is in fact a triple system, with the third, wider component
having a magnitude difference from the primary of over 5 magnitudes
in the 692-nm filter.

To give some feel for the basic properties of the sample of stars appearing
in Table 1 overall, we show in Figure 1(a) the magnitude difference as a 
function of the separation of the component from
the primary star. The majority of observations are clustered at very small
separations; these measurements 
would be difficult to obtain at smaller telescopes. The
dashed curve shown is an average-quality 5-$\sigma$ detection limit 
curve for DSSI at Gemini for the 692-nm filter. These curves are determined
by studying the statistics of local peaks in the reconstructed images we 
obtain; more about how it is calculated will be discussed in Section 3.2.
In Figure 1(b), we plot the magnitude difference observed as a function of
the total (system) apparent $V$ magnitude for the binary. This shows that
the stars we have observed so far for this project have magnitudes
in the range $6 < V < 10$, although it would be possible to observe 
much fainter sources at Gemini at high signal-to-noise; we plan to 
include fainter targets for this work in the future.

In Figure 2, we show contour plots of the reconstructed images obtained for
the triple system HIP 111805. The secondary star is itself a 
double-lined spectroscopic binary of period 551 days 
that has been sporadically detected
at the diffraction limit of the 6-m Special Astrophysical Observatory 
telescope by Balega \ea (2002, 2006, 2007). 
Panels (a) and (b) of the figure show the images at 692 nm and 880 nm
respectively obtained in 2013 July, and panels (c) and (d) show the same
for the 2014 July data. The asymmetric 
elongation of the secondary reveals that
it is in fact a binary of modest magnitude difference itself and 
that it has separation
near the diffraction limit of the telescope.
The position angle of this system has changed by nearly 
180$^{\circ}$ between the two observation epochs. This system will be
discussed further in Section 4.

\subsection{Relative Astrometry and Photometry}

To characterize the precision of the relative astrometry, we first
compared the results obtained in the two channels of the instrument
for the same observation by forming the differences between the two
channels for position angle and separation. These are shown in Figure 3. 
Considering only observations with separations from 0.0215 to 1.0 arc seconds,
we obtain an average difference in position angle of $0.24 \pm 0.37$ degrees.
For the separation values, the average difference is $-0.34 \pm 0.33$ mas. These 
values indicate that there is no measurable systematic error in the scale
or orientation values applied to the data. The standard deviation of the 
position angle differences is $1.99 \pm 0.26$ degrees, while for separation, 
we obtained a value of $1.79 \pm 0.24$ mas. Since we are forming a difference
between two measures of presumably the same uncertainty, these values
will be $\sqrt{2}$ larger than the intrinsic repeatability of an individual
measure. Therefore, we judge that, on average, the values in Table 1
have an intrinsic precision of approximately $1.41 \pm 0.18$ degrees
in position angle and $1.27 \pm 0.17$ mas in separation. These numbers would
be reduced by another factor of $\sqrt{2}$ by averaging the astrometric
results in both channels. While this was not done in Table 1, if one did
take that step, the values would be reduced to $1.00 \pm 0.13$ degrees
and $0.90 \pm 0.12$ mas. These are very much in line with the values 
obtained for the earlier observations published in Horch \ea (2012). 
The precision of the position angle is a function of separation, and 
degrades as the linear scatter subtends a larger
angle and the separation becomes smaller. Our measures have median
separation of $\sim$0.1 arcsec, so that the angular uncertainty 
will be dominated by the half of our measures below this, 
down to the diffraction limit. Taking an average separation about 0.05
arc seconds for these objects, we would expect them to have an angular
uncertainty of 
 $\arctan(0.00127/0.05) = 1.5^{\circ}$, based on the linear uncertainty
value. This is consistent with the angular uncertainty derived above.

Some measures in Table 1 have separations below the diffraction
limit. We have discussed this type of situation in Horch \ea (2006a) 
and Horch \ea (2011b), where we find in the latter reference that
comparing the results in the two channels of the instrument allows us
the ability to distinguish between elongation of speckles due to 
residual atmospheric dispersion and that due to the presence of an 
unresolved companion. For the measures below the diffraction limit
in Table 1, the consistency in the separation determination between
both channels of the instrument gives good that we are indeed measuring
the presence of an unresolved companion. Because the speckles from 
the primary and secondary stars are cases are blended in these cases,
there is some loss of precision in the measures we obtain; for example,
at WIYN, the uncertainty in separation roughly doubled for pairs 
observed below the diffraction limit with DSSI. While we do not yet
have enough measurements to characterize this at Gemini, it would seem
to be a reasonable assumption that the same is true at the larger aperture.

A number of systems in Table 1 have a previous orbit determination listed
in the Sixth Orbit Catalog (Hartkopf \ea 2001a). 
We have computed the ephemeris positions of 
these systems for the epoch(s) of observation shown in Table 1, and 
compared the separations and position angles to what we obtained. These
results are shown in Figure 4. In some cases, the orbital elements are 
published with uncertainties; in these cases, we can compute uncertainties
in the ephemeris positions themselves, and where ever possible, these have
been included in the figures. 

Especially in the separation plot, there are
three data points that deviate significantly from the zero line. These are
STF 1728 (at ephemeris 
separation 0.1358 arcsec) and the two observations of HO 295 
(at separations of 0.2701 and 0.3045 arcsec respectively). It is not clear at 
this point why the deviation of STF 1728 is so great. The orbit is relatively
recent and of excellent quality (Muterspaugh \ea 2010). However, the next
periastron passage is predicted to be in April of 2015, and the motion
is already relatively fast at this point. We have recomputed the orbit based on 
data from 1994 to the present (and including the points in Table 1), and 
we find that the data are consistent with a slightly shorter period (25.84
years versus 25.97, and a slightly different time of periastron passage
(2015.11 versus 2015.31), and with all other orbital elements similar to
the Muterspaugh \ea orbit. We suggest that if further observations of this
system can be made over the next year as the system goes through periastron,
then it may be sensible to revise the orbit at that point.
The deviation of the HO 295 points is explainable considering that this
is a triple system (as shown in Figure 2), and the current orbit for 
the AB pair is now almost 20 years old (Hartkopf \ea 1996).
In the next section, we present new orbital elements for this system.
Taking these exceptions into account, the data overall suggest that,
once again,
there is no evidence for a systematic error in the scale and orientation.

Turning now to the photometry, our previous papers (e.g. Horch \ea 2011a
and references therein) have discussed the
importance of establishing the ratio of the separation to the size
of the isoplanatic patch in order to have confidence in the differential
photometry obtained in speckle observations. As the isoplanatic patch
is inversely proportional to the seeing, a proxy parameter which we 
have called $q^{\prime}$ can be established as the seeing value times the 
separation of the pair. In general, the magnitude difference would 
be expected to be close to the true value for low values of $q^{\prime}$, and
as $q^{\prime}$ increases, then the $\Delta m$ obtained will be systematically
too large, as the decorrelation between the primary and secondary 
speckle patterns results in a loss of photon correlations at the 
expected separation.

In Figure 5(a), we plot the magnitude difference
we obtain here minus an average value obtained from previous 
measures appearing in the literature. Specifically, 
we examined all of the magnitude differences for these systems that
exist in the Fourth Interferometric Catalog of Hartkopf \ea (2001b),
and we select only those objects with three or more measures that 
were obtained with a filter within
20 nm of 692 nm. These are overwhelmingly dominated by our own measures
from the WIYN telescope, which are calibrated in a similar way, and 
adaptive optics measures of the CHARA group (ten Brummelaar \ea 1996). 
After removing YSC 134 from consideration as most of its previous measures were 
obtained below the diffraction limit at WIYN and will therefore
be somewhat uncertain, we are left with eleven 
comparison observations.
The result is that, while
the diagram is sparsely populated, there is excellent agreement 
between the literature values and the values in Table 1 for 
values of $q^{\prime}$ below 0.6. The two points above $q^{\prime} = 0.6$
have a larger observed $\Delta m$ than the value appearing in the 
literature, which is consistent with the loss of correlations due to
non-isoplanicity.
The mean difference below $q^{\prime}=0.6$ is $0.00 \pm 0.04$
magnitudes, and the standard deviation of these differences is $0.13 \pm 0.03$
magnitudes. Some of this scatter is due to the uncertainty of the 
literature values themselves; on average, the uncertainty is about 0.07 
magnitudes. If we subtract this in quadrature from 0.13, we obtain an
intrinsic repeatability in the magnitude difference of our measures of about 
0.11 magnitudes. The comparisons in the plot with {\it Hipparcos} values
(represented by the open circles) show a slight negative 
trend; this is expected due
to the fact that the {\it Hipparcos} filter is considerably bluer than 
692 nm. Most of the systems in question are known to be
main sequence systems, meaning that 
the secondary will be redder than the primary, and
therefore the system will exhibit a larger magnitude difference in a 
bluer filter.
Figure 5(b) shows a plot of the literature
values versus our values for those measures with $q^{\prime} < 0.6$; 
this should
be a line of slope one that passes through the origin, and the data
are consistent with that from 0 to 3 magnitudes.

\subsection{Non-detections}

Six systems that we observed showed no evidence of a companion to the
limit of detection. We show the detection limit in magnitude difference
from the primary for both 0.1 and 0.2 arc seconds in Table 2. As mentioned
earlier, in order to make a determination about the detection limit, we
use the reconstructed images obtained from the speckle data reduction. In
these images, the primary star is always centered in the image. We investigate
the image properties within concentric annuli centered on the primary. Within
each annulus, we determine the value of all local maxima. The background level
is then set to the average value of these maxima, and a standard deviation
of the peak values is computed. The detection limit for that particular annulus
is set as the background value plus five standard deviations, that is, it 
is a 5-$\sigma$ limit. A similar calculation is performed on the absolute value of
the local minima to make sure that the distributions of maxima and minima are similar. 
This is then associated with the mean radius of the
annulus. Once values for annuli with a range of different separations from the
primary have been computed, a cubic spline interpolation is performed to 
derive the detection limit curve as a function of separation. Generally, these
curves have very shallow values of the limiting $\Delta m$ at the smallest
separations, a rapid rise leading to a ``knee'' in the curve at approximately
0.1 arcsec, and a continued slower rise in limiting magnitude out to the
largest separations we measure (1.2 arcsec).

Examples of the detection limit curves for
one of the objects in Table 2 are shown in Figure 6. This object is 
actually a known
single-lined spectroscopic binary star with a 10-day period, but
based on the period, the spectral type of K2.5V, and the system parallax
using the revised {\it Hipparcos} value of $42.13 \pm 0.68$ mas (van Leeuwen, 
2007), we can roughly
estimate that the semi-major axis of the orbit is on the order of a few mas;
this would not be detectable using DSSI at Gemini even if the magnitude
difference were small. The curves show that there is no other
other wider component seen in the vicinity of this star. The other objects in 
Table 2 had no previous detection of a component either via spectroscopy or
high-resolution imaging.

\section{New Orbital Elements}

From the objects listed in Table 1, we have selected 9 for which the
addition of Gemini data permits either an orbit revision or a first 
orbit determination. All of these systems are spectroscopic binaries,
and so, recognizing that in most cases we do not have sufficient astrometry
to calculate a good-quality visual orbit, we have fixed the values of
period ($P$), time of periastron passage ($T_{0}$), eccentricity ($e$), 
and position angle of
the node in the plane of the true orbit ($\omega$) to be those of the 
spectroscopic orbit prior to fitting for the other three elements, namely
semi-major axis ($a$), inclination ($i$), and the node ($\Omega$). The 
orbit code of MacKnight \& Horch (2004) was then used to determine these three
elements and their uncertainties. This code is a grid search of orbital
parameters between user-chosen minimum and maximum values, followed by a 
downhill simplex algorithm to fine-tune the final result. We show our
results for the nine systems on which we used this approach in Tables 3 and
4. Table 3 gives elements for systems that already have an orbit determination
in the literature, whereas Table 4 gives the elements for the first
orbit determinations. One exception to the method regarding the use of 
the spectroscopic orbital elements is BAG 15Ba,Bb. In this case, fixing the
spectroscopic elements resulted in an orbit with large residuals; we then
determined this orbit solely from the relative astrometry, which produced
much lower residuals. In studying the difference in the orbital parameters,
the time of periastron passage was the only element that was significantly 
different from those of the spectroscopic orbit. It is not possible to assess
whether the value we obtained is consistent with that of the spectroscopic 
orbit as no uncertainties were given in that case (Duquennoy 1987).
In Figure 7, we show four sample orbits of the systems in Tables 3 and 4.

\subsection{Comments on Individual Systems}

\subsubsection{HSL 1}

As discussed in Horch \ea (2006b), this system is a metal-poor hierarchical
quadruple system where only the inner three stars have been imaged in the
current work due to the very small field of view of the DSSI at Gemini.
While data in the literature for the third component
of this nearly--edge-on system
indicate substantial motion over the last decade
that appears highly likely to be orbital in nature, obtaining a definitive orbit
will be a difficult proposition until we have almost a full orbit.
With less than half the orbit traced out, the period remains uncertain on 
the level of years, and there is also substantial uncertainty in the
semi-major axis.

The Gemini data do however help with the inner pair: they are much
more consistent with the visual+spectroscopic orbit in Horch \ea (2006b) than
the second orbit calculation there, which is
an unconstrained visual orbit that does not use the 
spectroscopic information. This confirms that the system has an inclination
near 90$^{\circ}$.
Since the inner pair is a double-lined spectroscopic binary, 
we can independently determine the parallax by comparing the $a \sin i$ value
from the spectroscopic orbit (in units of Gm) with the value of $a \sin i$ 
implied from the visual orbital elements (in arc seconds), 
and in doing this we obtain
$22.9 \pm 0.5$ mas. Horch \ea 2006b computed a value of $23.7 \pm 0.7$, which
is consistent with the value here. On the other hand, the revised 
{\it Hipparcos}
value of $19.76 \pm 0.82$ mas is almost 4$\sigma$ away from our
results; the motion of the third component, which went undetected by 
{\it Hipparcos},
may have some role in this discrepancy.

\subsubsection{YSC 134}

This K2.5V system has [m/H] $=-0.80$, and is a double-lined system. A
visual orbit has recently been computed by Docobo \ea (2013). Their method,
like the work here, incorporates the spectroscopic orbital elements, 
and the orbit is very similar to the one we obtain, although they did
not have all of the Gemini data points presented here. Calculating the implied 
parallax, we obtain $34.9 \pm 6.7$ mas, which is consistent with the 
{\it Hipparcos}
revised result of $39.98 \pm 0.73$ mas, within the uncertainty.

\subsubsection{A 1470}

A slightly metal-poor system with [m/H] of $-0.11$, the spectral
type appearing in SIMBAD is that of a G0 subgiant. 
Using the
spectroscopic and visual orbit data, a parallax of $14.07 \pm 0.22$ mas is 
obtained, which agrees very well with the {\it Hipparcos} revised result of 
$14.15 \pm 0.74$ mas. From this, an absolute V magnitude of 3.94 is obtained.
Although the $B-V$ color of the system is consistent with the SIMBAD
spectral assignment, the absolute magnitude for the system would seem to be too
faint for a subgiant, and more consistent with two slightly later dwarfs.
For this reason, we will treat the system as such in the next section.

\subsubsection{HO 295AB and BAG 15Ba,Bb}

This is the triple system appearing in Figure 2 and it has [m/H] of -0.29. 
The wider component, 
HO 295AB, has a 30 year period, but is nonetheless known as a 
spectroscopic double-lined system. Both this component and the 
smaller-separation component (an SB2 with a 1.5-year period) are used 
in the study presented in the next section. 
Our values of parallax from both orbits are 
reasonably consistent
with {\it Hipparcos}: for the inner pair, we have $25.5 \pm 0.5$ mas and for the
wider component, we obtain 24.1 mas, though an uncertainty estimate is not
possible in this case because the spectroscopic orbit was published without
error estimates. The revised {\it Hipparcos} parallax is $26.18 \pm 0.60$ mas.
Both orbits obtained here are shown in Figure 7.

\subsubsection{DSG 7Aa,Ab}

The measures in Table 1 represent the first resolution of the companion
to date. A K0 SB1 with very low metal abundance 
([m/H]=-1.39), this pair has a period of 226 days and a magnitude
difference of approximately 1.5. 
Ideally, one would like to have observations at several
different epochs spread out around the orbit, but we have just two 
observations, both taken in July 2013. Nonetheless, our measures are 
consistent with the astrometric orbit in the {\it Hipparcos} Catalogue
(ESA 1997) in terms of their position angle at the time of observation, 
and they allow us to determine the
size of the orbit to about 13\%. 

\subsubsection{YSC 132Aa,Ab}

Observations with DSSI at WIYN indicated a companion to this star that was
below the diffraction limit of that telescope. The first resolved
image came from our July 2012 Gemini run. In July 2013, the system 
had moved considerably, and by July of 2014, it was possible to 
determine the orbit with high reliability, in combination with the
spectroscopic information. The orbit of this system is shown in Figure 7.

\subsubsection{DSG 6Aa,Ab}

This triple system consists of an inner pair which is a single-lined
spectroscopic orbit, and wider component first resolved in 1999 by Mason
\ea (2001b). Subsequent observations have shown that this wider component
does orbit the inner pair, and in fact must have a period of 
$\sim$20 years. We have made a preliminary orbit calculation that is not presented
here; the period and semi-major axis uncertainties led to a large 
uncertainty in mass so that the system would not yet be useful in 
the study that we present in the next section. Interestingly,
the inclination 
of that calculation would seem to suggest that the orbit is not
coplanar with the spectroscopic orbit, so a full analysis of the 
system as a whole is warranted. There is 
archival data from the Fine Guidance Sensors on the {\it Hubble} Space 
Telescope that could be incorporated into a more detailed analysis
of both orbits, as well as unpublished WIYN speckle data. However,
that effort is 
beyond the scope of the present paper.

\subsubsection{DSG 8}

The measures in Table 1 are the first resolution of this well-known
double-lined spectroscopic binary with period 3.2 years. 
The observations are separated by nearly 180 degrees, allowing for 
a very good orbit when combining with the spectroscopic orbital
parameters. The orbit is shown in Figure 7. This immediately shows
why there have not been previous measures: the system is below
the diffraction limit even of Gemini for a large portion of the
orbit, due to the inclination. While the spectral type in the literature
is F2IV-V, the system is slightly redder than F2; we treat this 
system as an F5+F5.5 dwarf pair in the next section. The parallax
that we can obtain from our SB+visual orbit is in
complete agreement with the revised {\it Hipparcos} value: $14.3 \pm 0.3$ mas
in our case versus $14.51 \pm 0.47$ mas from {\it Hipparcos}.

\subsection{Total Mass as a Function of Metallicity}

Table 5 lists some further observed properties of the nine systems 
identified above. The columns give (1) the discoverer designation;
(2) the {\it Hipparcos} number; (3) the revised {\it Hipparcos} parallax of 
Van Leeuwen (2007); (4) the absolute magnitude obtained from the 
apparent magnitude (not listed) and the parallax, where no reddening
correction was made since these systems are all nearby; (5) the 
(composite) spectral type as it appears in SIMBAD; the metal abundance from 
Holmberg \ea (2009), unless the value
is otherwise marked; (6) the $B-V$ value listed in the {\it Hipparcos}
Catalogue, and (7) the average magnitude difference at 692 nm from 
all available DSSI measures of the target. Note that the absolute
magnitudes show clearly that all nine systems are on or very near
the main sequence.

Using the most recent Yale isochrones (Spada \ea 2013), 
we investigated the behavior of stellar mass as a function of
metal abundance at fixed $B-V$ color. 
We selected main sequence stars that had spectral types from 
mid-F to mid-K; this range is similar to the nine systems for which we 
obtained orbital elements (and hence total masses). The ages we chose 
were from 0.1 to 4 Gyr; this range insures that the spectral types in question will
all be close to the main sequence, which we know is the case for all of the
systems under study. For a given metal abundance and ranging from $B-V=0.44$
to 1.15, we then calculated the 
ratio of the mass extracted from the isochrones to the mass
for a star of the same color but with solar metallicity. By definition,
this function is one when [m/H]=0.0, but the isochrones predict that
as the metallicity decreases, the ratio also decreases to a value of 
approximately 0.6 at [m/H]=-1.5. While there
is some variation depending on age and the choice of the mixing length parameter,
we found that there was little dependence on a star's color (or equivalently,
spectral type) over the range of interest here, as shown in Figure 8. 
Since the curves are nearly independent of spectral type, it should be 
true that the total mass of a binary star will follow the same trend,
provided that both stars fall in the spectral range of the simulations. We also 
found the same result with the older Yale-Yonsei isochrones found in 
Demarque \ea (2004).

In Tables 3 and 4, we have dynamical estimates of the total mass of 9 systems 
that span a considerable range in metal abundance. If we could estimate
the mass of the solar-abundance analog for the system, then we could 
compute the ratio and examine whether the trend is similar to what the
isochrones predict. To make this mass estimate, we use the composite $B-V$
color for the system as it appears in the {\it Hipparcos} Catalogue
and the average magnitude
difference measured by DSSI at 692 nm, combining all observations of the 
system in that filter to date at Gemini, 
{\it i.e.\ }the last two columns of Table 5.
We then use the solar-abundance spectral library of Pickles (1998) to 
combine stars of different spectral types to produce a composite $B-V$ value
and $\Delta m$ at 692 nm that is as close as possible to the measured values.
We incorporate into these models a standard atmospheric transmission curve,
the known filter transmission curve, and dichroic transmission curve
for DSSI. We consider only the
692- and not the 880-nm data at this stage because we have a more reliable 
transmission curve for this filter. After giving the same
two columns to identify the objects as in Table 5, Table 6 shows 
in the third and fourth columns the
assigned component spectral types and the composite $B-V$ that would be
obtained. The latter should be directly comparable to the next-to-last column of
Table 5. Figure 9(a) shows that the scatter is modest; the standard deviation
of the difference between the measured and simulated colors is 
less than 0.03 magnitudes. Figure 9(b) shows a similar result for the
simulated $\Delta m$ values at 692 nm; here, the uncertainty is dominated
by the measured values and not the simulation.

We can then convert these assigned spectral types into mass estimates;
for this we used the standard reference of Schmidt-Kaler (1982), and
these appear in Columns 6 and 7 of Table 6. 
We have estimated the uncertainty of
the total masses shown by using the scatter in the 
$B-V$ color. Specifically, we determine a range of spectral types 
possible for each component within the color uncertainty, 
we read off the masses corresponding to these
high and low estimates of spectral type, and then 
use them to set the uncertainty
interval for the mass of the component. Although the uncertainties in
components are not shown in Table 6, we added those in quadrature to
obtain an uncertainty estimate in the implied total mass at solar abundance.
We also checked the conversion from spectral type to mass using the information
provided in the recent work of Boyajian \ea (2012a, 2012b), and found 
very good agreement with the Schmidt-Kaler
reference over the spectral range of interest.
Finally, in the last column of Table 6, we show the dynamical total mass
estimate using Kepler's harmonic law and the data in Tables 3 and 4. While it is
not used in the analysis here, it is worth noting that for the systems that are 
double-lined spectroscopic binaries, the mass ratios $m_{2}/m_{1}$ implied from
Table 6 are in reasonably good agreement from the values implied from the
spectroscopic orbits in all cases.

In Figure 10, we plot the theoretical and observed ratio of mass
to the solar-abundance mass at the same color, as a function of
metal abundance.
We have assumed an uncertainty in metallicity of the observed data of 0.1 dex
(which is the uncertainty stated in Holmberg \ea [2009], 
the source of most of our abundance values).
The plot suggests that, within the uncertainty, the points follow the 
trend expected from the stellar structure calculations. With further work
on these and potentially other systems yet to be identified, 
it should be possible to shrink the vertical error 
bars in the plot to make more definitive statements concerning the
agreement between the observational data and stellar models for a wide range of
metallicity. 

\section{Conclusions}

We have presented speckle results on a number of spectroscopic binaries
and other nearby stars taken from the Gemini North telescope with the 
DSSI dual-channel speckle imaging system. These systems are all nearby, 
but span a range of
metal abundance from near the solar value to [m/H]=-1.39. 
The precision of the astrometry 
appears to be in line with that of our previous work at Gemini, namely 
about 1 mas, and the differential photometry of the observations is 
precise at the 0.1 magnitude level.

We have used the information presented here together with other astrometry
in the literature and the known spectroscopic orbits to obtain high-quality
values for the semi-major axis, inclination, 
and ascending node for 9 systems. Using the
semi-major axes together with the spectroscopic
periods and known parallaxes
resulted in total mass estimates. These were compared with
theoretical predictions for F through M main sequence stars as a function
of metallicity. We find that, over this spectral range, our results are so far 
consistent with the predicted overall
trend toward a lower mass at a given spectral type as the 
metal abundance decreases. 

\acknowledgments
We thank the {\it Kepler} 
Science Office located at the NASA Ames Research Center
for providing partial financial support for the upgraded DSSI instrument.
It is also a pleasure to thank Steve Hardash, Andy Adamson, Inger Jorgensen,
John White,
and the entire summit crew for their excellent work in getting the instrument
to the telescope and installing it.
This work was funded by the {\it Kepler} Science Office and NSF Grant
AST-1429015.
It made use of the Washington
Double Star Catalog maintained at the U.S. Naval Observatory, the
SIMBAD database, operated at CDS, Strasbourg, France, and the 
9th Catalog of Spectroscopic Orbits of Binary Stars.




\begin{centering}

\pagestyle{empty}
\begin{deluxetable}{lllrlrrrr}
\tabletypesize{\scriptsize}
\tablenum{1}
\tablecaption{Binary star speckle measures.}
\tablehead{ 
\colhead{WDS} & 
\colhead{HR,ADS} & 
\colhead{Discoverer} & HIP & 
\colhead{Date} & 
 \colhead{$\theta$} & \colhead{$\rho$} & 
\colhead{$\Delta m$} & 
\colhead{$\lambda$} \\
\colhead{($\alpha$,$\delta$ J2000.0)} & 
\colhead{DM,etc.} & \colhead{Designation} && 
\colhead{(Bess. Yr.)} & 
\colhead{($^{\circ}$)} & 
\colhead{(${\prime \prime}$)} 
 & 
 & \colhead{(nm)}  
} 
\startdata 

$00022+2705$ & ADS 17175 & BU 733AB &     171 &
   2014.5619 & 339.2 & 0.4016 & 2.64 & 692 \\
 &&&&
   2014.5619 & 339.3 & 0.4024 & 2.21 & 880 \\

$00063+5826$ & HD    123 & STF 3062AB & 518 &
   2013.5734 & 353.7 & 1.5411 &  ... & 692\tablenotemark{a} \\
 &&&&
   2013.5734 & 353.8 & 1.5395 &  ... & 880\tablenotemark{a} \\
 &&&&
   2014.5591 & 355.5 & 1.5479 &  ... & 692\tablenotemark{a} \\
 &&&&
   2014.5591 & 355.6 & 1.5552 &  ... & 880\tablenotemark{a} \\

$00089+2050$ & G 131-26 & BEU 1 &   ...   &  
   2014.5646 &  94.5 & 0.1458 & 0.77 & 692 \\
 &&&&
   2014.5646 &  94.8 & 0.1460 & 0.43 & 880 \\

$00133+6920$ & GJ 11 & KUI 1 &   1068  &  
   2014.5646 &  96.4 & 0.8484 & 0.68 & 692 \\
 &&&&
   2014.5646 & 276.7 & 0.8492 & 0.38 & 880 \\

$00325+6714$ & ADS 440 & MCY 1Aa,Ab &   2552  &  
   2014.5646 & 222.2 & 0.3703 & 3.05 & 692 \\
 &&&&
   2014.5646 & 222.2 & 0.3694 & 2.32 & 880 \\

$02128-0224$ & HD 13612   & TOK 39Aa,Ab & 10305 &
   2013.5734 & 142.3 & 0.0194 & 0.29 & 692 \\
 &&&&
   2013.5734 & 140.0 & 0.0205 & 0.31 & 880 \\
 &&&&
   2014.5646 & 127.2 & 0.0193 & 0.51 & 692 \\
 &&&&
   2014.5646 & 130.9 & 0.0206 & 0.42 & 880 \\

$02278+0426$ & HD 15285 & A 2329 &   11452 & 
   2014.5646 & 347.4 & 0.1442 & 0.09 & 692 \\
 &&&&
   2014.5646 & 347.5 & 0.1446 & 0.30 & 880 \\

$13100+1732$ & HD 114378 & STF 1728AB &   64241 & 
   2014.5607 &  12.2 & 0.0906 & 0.17 & 692 \\
 &&&&
   2014.5607 &  12.2 & 0.0908 & 0.26 & 880 \\

$14035+1047$ & HD 122742 & GJ 538 &   68682 & 
   2014.5636 &  76.5 & 0.4052 & 3.80 & 692 \\
 &&&&
   2014.5636 &  76.7 & 0.4047 & 2.99 & 880 \\

$14539+2333$ & GJ 568 & REU 2 &   72896 & 
   2014.5608 &  92.2 & 0.9591 & 1.36 & 692 \\
 &&&&
   2014.5608 &  92.5 & 0.9573 & 1.11 & 880 \\

$16329+0315$ & HD 149162 & DSG 7 Aa & 81023 &
   2013.5615 & 326.7 & 0.0174 & 1.18 & 692 \\
 &&&&
   2013.5615 & 321.9 & 0.0195 & 1.17 & 880 \\
 &&&&
   2013.5668 & 124.1 & 0.0069 & 1.37 & 692 \\
 &&&&
   2013.5668 & 124.1 & 0.0164 & 1.23 & 880 \\
$16329+0315$ & HD 149162 & DSG 7 Aa-B & 81023 &
   2013.5615 & 227.9 & 0.2824 & 5.63 & 692 \\
 &&&&
   2013.5615 & 226.5 & 0.2841 & 4.85 & 880 \\
 &&&&
   2013.5668 & 227.5 & 0.2881 & 5.33 & 692 \\
 &&&&
   2013.5668 & 228.5 & 0.2824 & 3.84 & 880 \\

$17080+3556$ & ADS 10360 & HU 1176AB & 83838 &
   2013.5642 &  ...  &   ...  & 0.30 & 692\tablenotemark{b} \\
 &&&&
   2013.5642 &  ...  &   ...  & 0.31 & 880\tablenotemark{b} \\
 &&&&
   2014.5471 &  ...  &   ...  & 0.40 & 692\tablenotemark{b} \\
 &&&&
   2014.5471 &  ...  &   ...  & 0.30 & 880\tablenotemark{b} \\

$17247+3802$ & BD+38 2932 & HSL   1Aa,Ab & 85209 &
   2013.5643 &  74.3 & 0.0050 & 0.58 & 692 \\
 &&&&
   2013.5643 &  74.3 & 0.0044 & 0.17 & 880 \\
 &&&&
   2014.5471 &  53.0 & 0.0233 & 0.13 & 692 \\
 &&&&
   2014.5471 &  51.9 & 0.0232 & 0.19 & 880 \\
$17247+3802$ & BD+38 2932 & HSL   1Aa,Ac & 85209 &
   2013.5643 &  61.3 & 0.1615 & 2.13 & 692 \\
 &&&&
   2013.5643 &  61.3 & 0.1649 & 1.71 & 880 \\
 &&&&
   2014.5471 &  59.4 & 0.2230 & 2.21 & 692 \\
 &&&&
   2014.5471 &  59.6 & 0.2226 & 1.92 & 880 \\
   
 \tablebreak

$18099+0307$ & ADS 11113 & YSC 132Aa,Ab & 89000 &
   2013.5643 &  35.1 & 0.0156 & 0.20 & 692 \\
 &&&&
   2013.5643 &  36.4 & 0.0149 & 0.01 & 880 \\
 &&&&
   2014.5636 &  92.6 & 0.0205 & 0.00 & 692 \\
 &&&&
   2014.5636 &  91.0 & 0.0204 & 0.29 & 880 \\

$19027+4307$ & HD 177412 & YSC 13 &   93511 & 
   2014.5640 & 247.9 & 0.0372 & 0.83 & 692 \\
 &&&&
   2014.5640 & 248.0 & 0.0382 & 0.90 & 880 \\

$19264+4928$ & GJ 1237 & YSC 134 & 95575 & 
   2013.5674 & 341.5 & 0.0209 & 0.25 & 692 \\
 &&&&
   2013.5674 & 159.1 & 0.0254 & 0.48 & 880 \\
 &&&&
   2014.5614 & 289.8 & 0.0222 & 0.39 & 692 \\
 &&&&
   2014.5614 & 281.2 & 0.0227 & 0.49 & 880 \\
 &&&&
   2014.5640 & 271.4 & 0.0237 & 0.96 & 692 \\
 &&&&
   2014.5640 & 272.2 & 0.0225 & 0.71 & 880 \\

$21041+0300$ & HD 200580 & WSI 6AB & 103987 &
   2013.5677 & 275.8 & 0.2494 & 1.92 & 692 \\
 &&&&
   2013.5677 & 275.9 & 0.2484 & 1.79 & 880 \\
 &&&&
   2014.5645 & 282.0 & 0.2561 & 1.84 & 692 \\
 &&&&
   2014.5645 & 282.3 & 0.2555 & 1.75 & 880 \\
$21041+0300$ & HD 200580 & DSG 6Aa,Ab & 103987 &
   2013.5677 & 242.0 & 0.0151 & 1.18 & 692 \\
 &&&&
   2013.5677 & 245.6 & 0.0188 & 1.77 & 880 \\
 &&&&
   2014.5645 & 216.1 & 0.0302 & 1.94 & 692 \\
 &&&&
   2014.5645 & 215.2 & 0.0346 & 2.05 & 880 \\

$21145+1000$ & HD 202275 & STT 535 & 104858 &
   2013.5704 &  ...  &   ...  & 0.28 & 692\tablenotemark{b} \\
 &&&&
   2013.5704 &  ...  &   ...  & 0.27 & 880\tablenotemark{b} \\
 &&&&
   2014.5616 &  ...  &   ...  & 0.19 & 692\tablenotemark{b} \\
 &&&&
   2014.5616 &  ...  &   ...  & 0.26 & 880\tablenotemark{b} \\

$21148+3803$ & HD 202444 & AGC 13AB &  104887 & 
   2014.5587 &  ...  &   ...  & 2.77 & 692\tablenotemark{b} \\
 &&&&
   2014.5587 &  ...  &   ...  & 2.64 & 880\tablenotemark{b} \\

$22357+5312$ & HD 214222 & A 1470 & 111528 &
   2013.5704 &  33.1 & 0.0757 & 0.24 & 692 \\
 &&&&
   2013.5704 &  33.2 & 0.0758 & 0.27 & 880 \\
 &&&&
   2014.5644 &  53.4 & 0.0928 & 0.00 & 692 \\
 &&&&
   2014.5644 &  53.4 & 0.0930 & 0.21 & 880 \\
 
$22388+4419$ & HD 214608 & HO 295AB & 111805 &
   2013.5704 & 333.7 & 0.2497 & 0.46 & 692 \\
&&&&
   2013.5704 & 333.5 & 0.2509 & 0.35 & 880 \\
&&&&
   2014.5644 & 334.3 & 0.3072 & 0.43 & 692 \\
&&&&
   2014.5644 & 334.6 & 0.3075 & 0.38 & 880 \\
$22388+4419$ & HD 214608 & BAG 15Ba,Bb & 111805 &
   2013.5704 & 334.9 & 0.0390 & 1.57 & 692 \\
&&&&
   2013.5704 & 332.4 & 0.0418 & 1.44 & 880 \\
&&&&
   2014.5644 & 150.9 & 0.0258 & 0.31 & 692 \\
&&&&
   2014.5644 & 155.9 & 0.0280 & 0.20 & 880 \\


$23347+3748$ & HD 221757 & YSC 139 & 116360 &
   2013.5704 &  93.8 & 0.0337 & 0.46 & 692 \\
&&&&
   2013.5704 &  93.7 & 0.0341 & 0.50 & 880 \\
&&&&
   2014.5618 &  93.5 & 0.0295 & 0.50 & 692 \\
&&&&
   2014.5618 &  92.4 & 0.0311 & 0.37 & 880 \\

\tablebreak

$23485+2539$ & HD 223323 & DSG 8 & 117415 & 
   2013.5680 & 293.5 & 0.0231 & 0.05 & 692 \\
&&&&
   2013.5680 & 293.3 & 0.0228 & 0.00 & 880 \\
&&&&
   2014.5618 & 300.5 & 0.0291 & 0.08 & 692 \\
&&&&
   2014.5618 & 300.4 & 0.0292 & 0.13 & 880 \\

\enddata 
 
\tablenotetext{a}{Photometry for this observation does not appear because
the $q^{\prime}$ factor discussed in the text was above 0.6 arcsec$^{2}$.}
\tablenotetext{b}{Astrometry for this observation does not appear because
it was used in the determination of the scale.}

\end{deluxetable} 


\begin{deluxetable}{ccccccc}
\tabletypesize{\scriptsize}
\tablewidth{0pt}
\tablenum{2}
\tablecaption{High-Quality Non-detections and 5-$\sigma$ Detection Limits}
\tablehead{
\colhead{($\alpha$,$\delta$ J2000.0)} & {\it Hipparcos} &
\colhead{Date} &
\multicolumn{2}{c}{5-$\sigma$ Det. Lim., 692 nm} &
\multicolumn{2}{c}{5-$\sigma$ Det. Lim., 880 nm} \\
\colhead{(WDS format)} & Number &
\colhead{(Bess. Yr.)} &
\colhead{0.1$^{\prime \prime}$} &
\colhead{0.2$^{\prime \prime}$} &
\colhead{0.1$^{\prime \prime}$} &
\colhead{0.2$^{\prime \prime}$} 
}
\startdata
01291+2143 &   6917 & 2014.5619 & 4.00 & 4.83 & 3.94 & 4.72 \\
14308+3527 &  70950 & 2014.5636 & 4.23 & 4.80 & 4.29 & 5.06 \\
16255+7123 &  80467 & 2013.5615 & 2.66 & 4.08 & 4.01 & 5.04 \\
16440+0901 &  81923 & 2014.5636 & 4.14 & 5.02 & 4.11 & 4.99 \\
22057+1223 & 109067 & 2013.5677 & 3.97 & 4.62 & 3.75 & 4.93 \\
22316+0210 & 111195 & 2013.5677 & 4.27 & 4.91 & 3.83 & 4.70 \\
\enddata
\end{deluxetable}

\clearpage

\begin{deluxetable}{lcccc}
\tabletypesize{\scriptsize}
\tablewidth{0pt}
\tablenum{3}
\tablecaption{Visual Orbit Refinements for Four Systems}
\tablehead{ 
\colhead{Parameter} &
\colhead{HSL 1Aa,Ab} &
\colhead{YSC 134} & 
\colhead{A 1470} & 
\colhead{HO 295AB} 
} 
\startdata 
HIP  & 85209 & 95575 & 111528 & 111805 \\
Type of Spectroscopic Orbit & SB2\tablenotemark{a} & SB2\tablenotemark{b} & SB2\tablenotemark{c} & SB2\tablenotemark{d}\\
$P$, years         & $1.2283$ & $0.45677$ & $22.3455$ & $29.9995$ \\
$a$, mas           & $31.6 \pm 0.6$ & $25.7 \pm 1.2$ & $144.6 \pm 0.8$ 
& $332.1 \pm 0.3$ \\
$i$, degrees       & $81.9 \pm 1.8$ & $141.2 \pm 9.2$ & $63.3 \pm 0.3$ 
& $88.2 \pm 0.1$ \\
$\Omega$, degrees  & $56.7 \pm 1.5$ & $ 29.1 \pm 10.3$ & $110.5 \pm 0.3$
& $154.7 \pm 0.1$ \\
$T_{0}$, years     & $1986.3731$ & $1990.0465$ & $1985.2460$ & $1979.8000$ \\ 
$e$                & $0.1634$ & $0.139$ & $0.362$ & $0.30$ \\
$\omega$, degrees  & $356.6$ & $ 58.8$ & $144.2$ & $ 81.5$  \\
\enddata 

\tablenotetext{a}{The spectroscopic elements are fixed to those of 
Goldberg \ea 2002.}
\tablenotetext{b}{The spectroscopic elements are fixed to those of 
Halbwachs \ea 2012.}
\tablenotetext{c}{The spectroscopic elements are fixed to those of 
Pourbaix 2012.}
\tablenotetext{d}{The spectroscopic elements are fixed to those of 
Duquennoy 1987.}

\end{deluxetable}

\begin{deluxetable}{lccccc}
\tabletypesize{\scriptsize}
\tablewidth{0pt}
\tablenum{4}
\tablecaption{Preliminary Visual Orbits for Five Systems}
\tablehead{ 
\colhead{Parameter} &
\colhead{DSG 7Aa,Ab} &
\colhead{YSC 132Aa,Ab} & 
\colhead{DSG 6Aa,Ab} & 
\colhead{BAG 15Ba,Bb} & 
\colhead{DSG 8} 
} 
\startdata 
HIP  & 81023 & 89000 & 103987 & 111805 & 117415 \\
Type of Spectroscopic Orbit & SB1\tablenotemark{a} & SB2\tablenotemark{b} & 
SB1\tablenotemark{c} & SB2\tablenotemark{d} & SB2\tablenotemark{e}\\
$P$, years         & $0.61897$ & $0.54634$ & $1.03441$ & $1.502 \pm 0.024$ & 
$3.21725$ \\
$a$, mas           & $14.8 \pm 2.0$ & $18.9 \pm 0.6$ & $21.6 \pm 0.6$ 
& $41.8 \pm 0.9$ & $40.3 \pm 1.2$ \\
$i$, degrees       & $112 \pm 26$ & $169 \pm 13$ & $ -0.3 \pm 0.2$ 
& $88.3 \pm 1.3$ & $86.3 \pm 1.8$ \\
$\Omega$, degrees  & $155 \pm 15$ & $ 46.6 \pm 2.2$ & $101 \pm 26$
& $334.7 \pm 0.9$ & $120.6 \pm 2.3$ \\
$T_{0}$, years     & $1988.4317$ & $1996.1450$ & $1986.5691$ & 
$1986.517 \pm 0.042$ & $2004.6148$ \\
$e$                & $0.3114$ & $0.302$ & $0.086$ & $0.02 \pm 0.02$ & $0.604$ \\
$\omega$, degrees  & $203.62$ & $ 86.1$ & $83$ & $349 \pm 9$ & $258.8$ \\
\enddata 

\tablenotetext{a}{The spectroscopic elements are fixed to those of 
Latham \ea 2002.}
\tablenotetext{b}{The spectroscopic elements are fixed to those of 
Griffin 1999.}
\tablenotetext{c}{The spectroscopic elements are fixed to those of 
Latham \ea 1992.}
\tablenotetext{d}{All elements are calculated from the speckle data, 
but a double-lined orbit exists due to Duquennoy 1987. This orbit
has similar elements to the orbit here except for the time of 
periastron passage.}
\tablenotetext{e}{The spectroscopic elements are fixed to those of 
Griffin 2007.}

\end{deluxetable}

\begin{deluxetable}{lrcccrcc}
\tabletypesize{\scriptsize}
\tablewidth{0pt}
\tablenum{5}
\tablecaption{Further Observed Properties for the Systems in Tables 3 and 4.}
\tablehead{ 
\colhead{Name} &
\colhead{HIP} &
\colhead{$\pi$} & 
\colhead{Abs.\ V} & 
\colhead{Spectral} & 
\colhead{[m/H]} &
\colhead{$B-V$} &
\colhead{$\Delta m$} \\
&&
\colhead{(mas)} & 
\colhead{Mag.} & 
\colhead{Type} & 
&& (692 nm)
} 
\startdata 
HSL 1 Aa,Ab   &  85209 & $19.76 \pm 0.82$ & 4.94 & G5    &  
-0.75\tablenotemark{a} &
$0.76 \pm 0.02$ \tablenotemark{b} & $0.36 \pm 0.13$ \\
YSC 134       &  95575 & $39.98 \pm 0.73$ & 6.02 & K2.5V &  
-0.80 &
$0.929 \pm 0.009$ & $0.61 \pm 0.13$ \\
A 1470        & 111528 & $14.15 \pm 0.74$ & 3.94 & G0IV  &  
-0.11 &
$0.610 \pm 0.015$ & $0.12 \pm 0.12$ \\
HO 295AB      & 111805 & $26.18 \pm 0.60$ & 3.91 & G0    & 
-0.29 &
$0.581 \pm 0.005$ & $0.45 \pm 0.02$ \\
DSG 7Aa,Ab    &  81023 & $23.14 \pm 1.02$ & 5.64 & K0    & 
-1.39\tablenotemark{a}  &
$0.868 \pm 0.004$ & $1.28 \pm 0.10$ \\
YSC 132 Aa,Ab &  89000 & $21.31 \pm 0.31$ & 2.31 & F5V   & 
-0.13  &
$0.490 \pm 0.005$ & $0.07 \pm 0.07$ \\
DSG 6Aa,Ab    & 103987 & $19.27 \pm 0.99$ & 3.73 & F9V   & 
-0.51  &
$0.547 \pm 0.007$ & $1.54 \pm 0.22$ \\
BAG 15Ba,Bb   & 111805 & $26.18 \pm 0.60$ & 3.91 & G0    & 
-0.29 &
$0.581 \pm 0.005$ & $0.94 \pm 0.63$ \\
DSG 8         & 117415 & $14.51 \pm 0.47$ & 2.89 & F2IV-V & 
-0.46 &
$0.443 \pm 0.009$ & $0.07 \pm 0.02$ \\
\enddata 
\tablenotetext{a}{From Latham \ea 1992.}
\tablenotetext{b}{The {\it Hipparcos} $B-V$ has a large uncertainty; we use the
value shown in Horch \ea 2006b here.}
\end{deluxetable}

\begin{deluxetable}{lrlcclccc}
\tabletypesize{\scriptsize}
\tablewidth{0pt}
\tablenum{6}
\tablecaption{Mass Comparison for the Systems in Tables 3 and 4.}
\tablehead{ 
\colhead{Name} &
\colhead{HIP} &
\colhead{Assigned Component} &
\colhead{Derived} &
\colhead{Derived} &
\colhead{Derived} &
\colhead{Implied Total} &
\colhead{Total Mass} \\
&&
\colhead{Spectral Types} &
\colhead{$B-V$} &
\colhead{$\Delta m$} &
\colhead{Masses} &
\colhead{Mass} &
\colhead{from Orbit } \\
&&&&
\colhead{(692nm)} &
\colhead{($M_{\odot}$)\tablenotemark{a}} &
\colhead{($M_{\odot}$)\tablenotemark{a}} &
\colhead{($M_{\odot}$)}

} 
\startdata 
HSL 1 Aa,Ab   &  85209 & G5V,G8V(,K6V) & 0.71 & 0.34 & 
0.92, 0.84 & $1.76 \pm 0.07$ & 
$1.58 \pm 0.17$ \\
YSC 134       &  95575 & K2V,K4.5V & 0.96 & 0.60 &
0.74, 0.68 & $1.42 \pm 0.02$ &
$1.27 \pm 0.19$ \\
A 1470        & 111528 & G2V,G3V & 0.62 & 0.13 &
1.00, 0.97 & $1.97 \pm 0.08$ &
$2.14 \pm 0.34$  \\
HO 295AB      & 111805 & F9V,G5V+K1V & 0.59    &  0.45 &
1.12, 0.92+0.77 & 
$2.81 \pm 0.14$ &
$2.27 \pm 0.16$  \\
DSG 7Aa,Ab    &  81023 & K1V,K6V(,M5V) & 0.85 & 1.30 &
0.77, 0.64 & $1.41 \pm 0.03$ &
$0.69 \pm 0.32$  \\
YSC 132 Aa,Ab &  89000 & F7V,F7.5V & 0.49 & 0.08 &
1.26, 1.23 & $2.49 \pm 0.10$ &
$2.33 \pm 0.26$ \\
DSG 6Aa,Ab    & 103987 & F9V,G9V(,K0V) & 0.58   &  1.36 &
1.12, 0.82 & 
$1.94 \pm 0.08$ &
$1.32 \pm 0.23$ \\
BAG 15Ba,Bb   & 111805 & G5V,K1V & 0.59 & 0.92 &
0.92, 0.77 & $1.69 \pm 0.12$ &
$1.80 \pm 0.17$ \\
DSG 8         & 117415 & F5V,F5.5V & 0.44 & 0.06 &
1.40, 1.37 & $2.77 \pm 0.09$ &
$2.07 \pm 0.28$ \\
\enddata 

\tablenotetext{a}{These columns assume the Solar metal abundance.}
\end{deluxetable}

\end{centering}

\clearpage

\begin{figure}[tb]
\plottwo{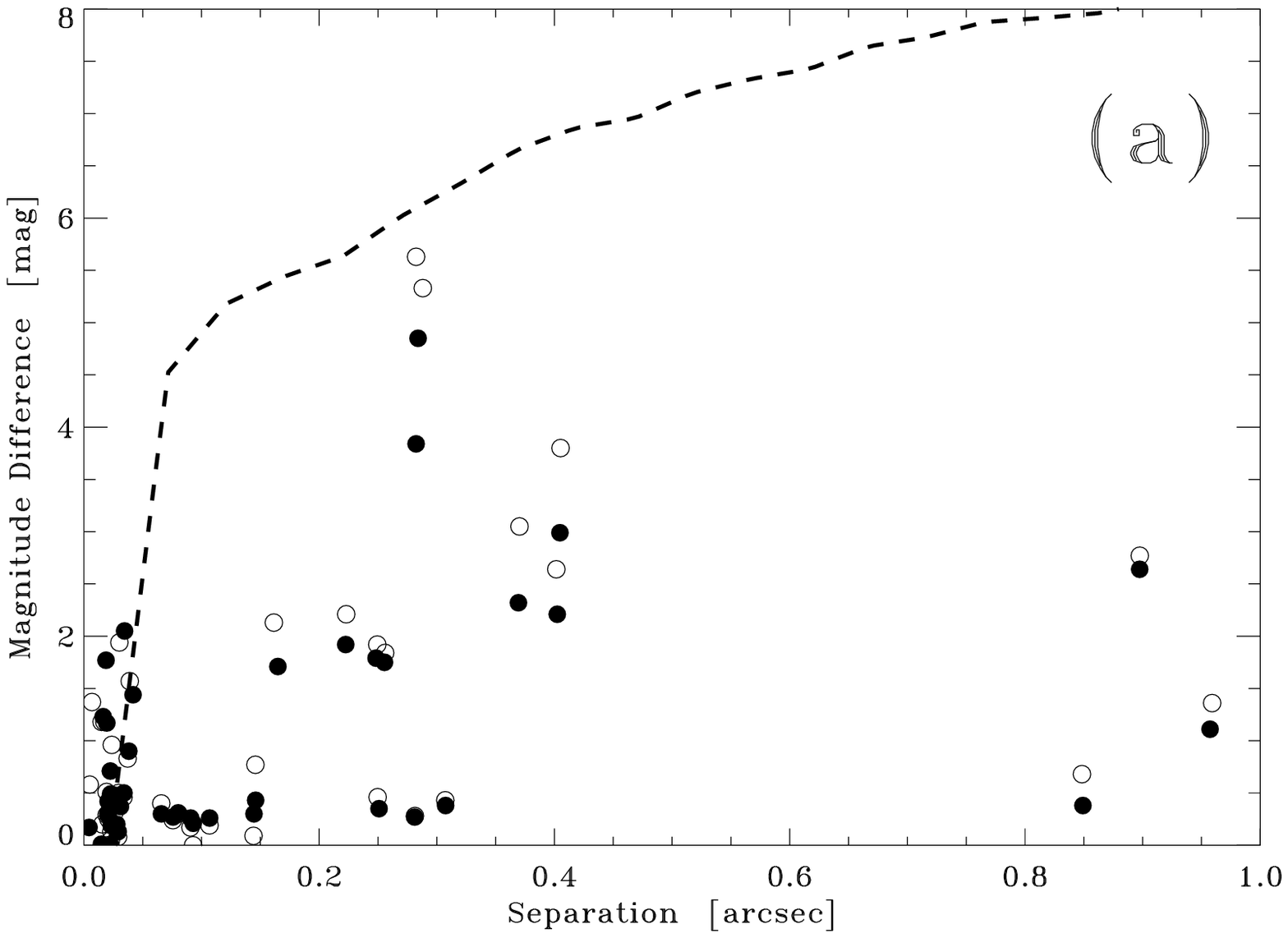}{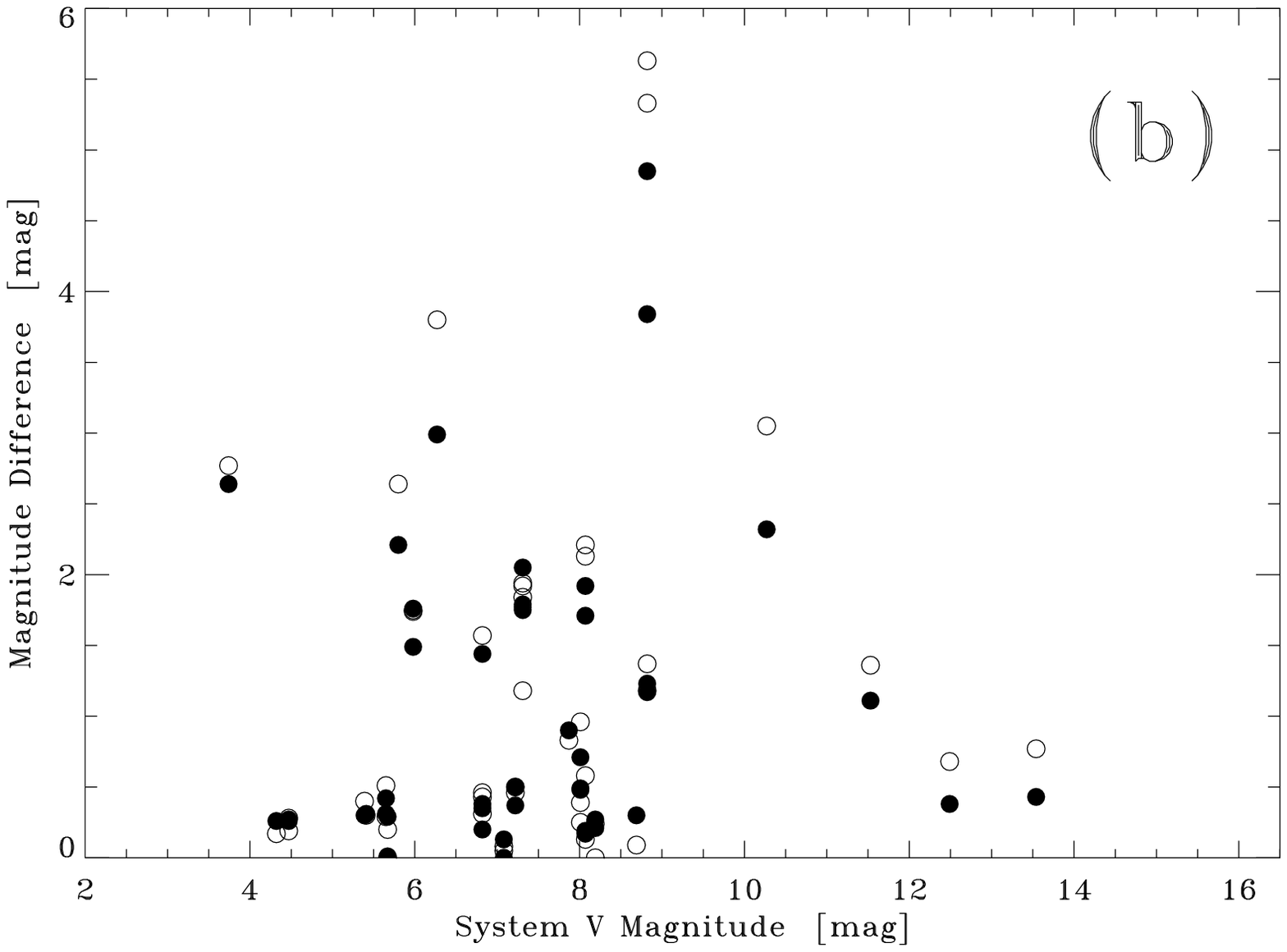}
\caption{
(a) Magnitude difference as a function of separation for the measures
listed in Table 1. The dashed curve is a typical 5-$\sigma$ detection limit
curve for the speckle camera at Gemini, such that above the curve, we would
not expect to make a definitive detection of a companion. (These curves are discussed
further in Section 3.2.)
(b) Magnitude difference as a function of system
$V$ magnitude for the measures listed in Table 1.
In both plots, the filled circles are measures taken with the 880 nm filter
and open circles are measures in the 692 nm filter. 
}
\end{figure}

\begin{figure}[tb]
\plottwo{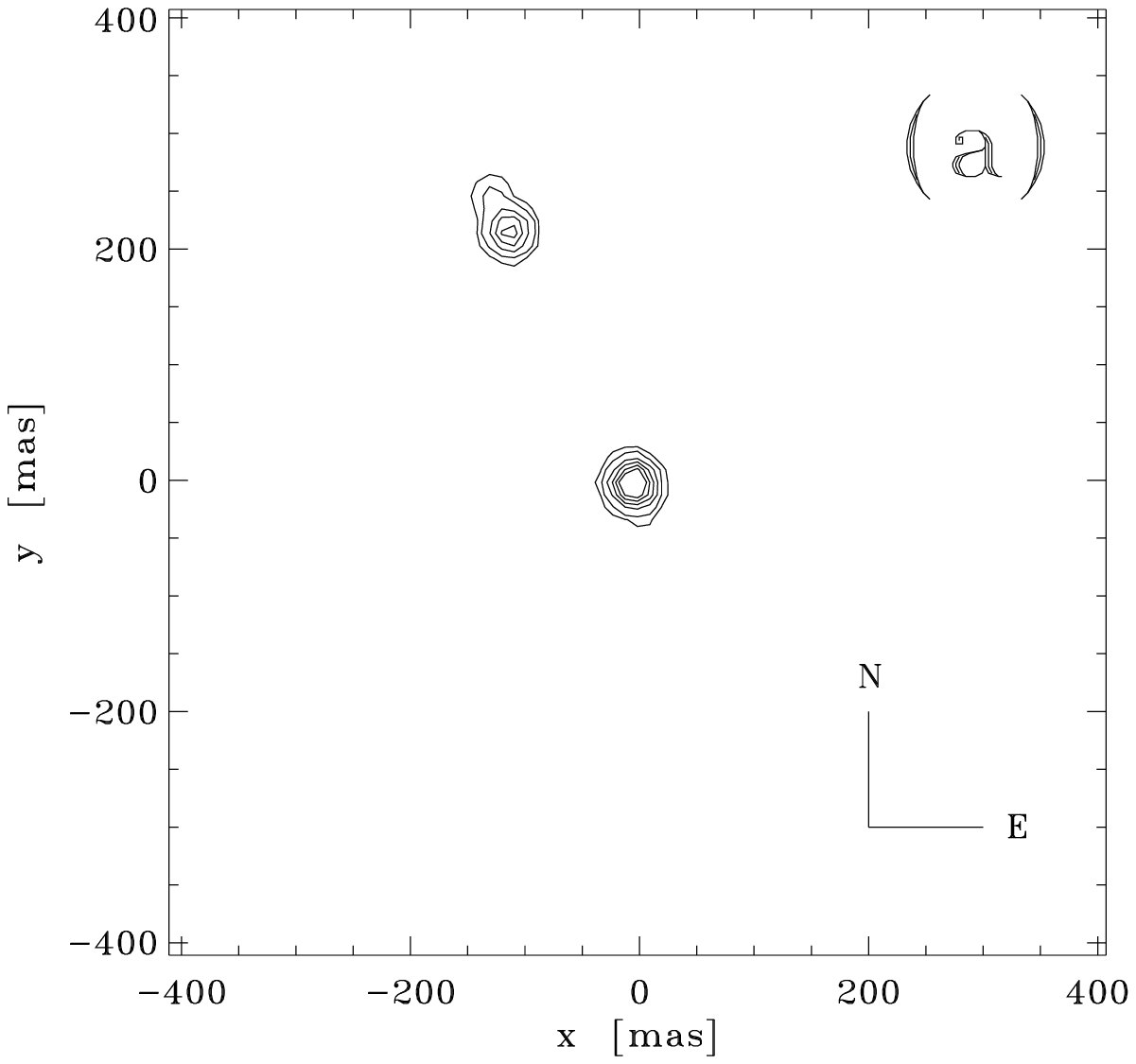}{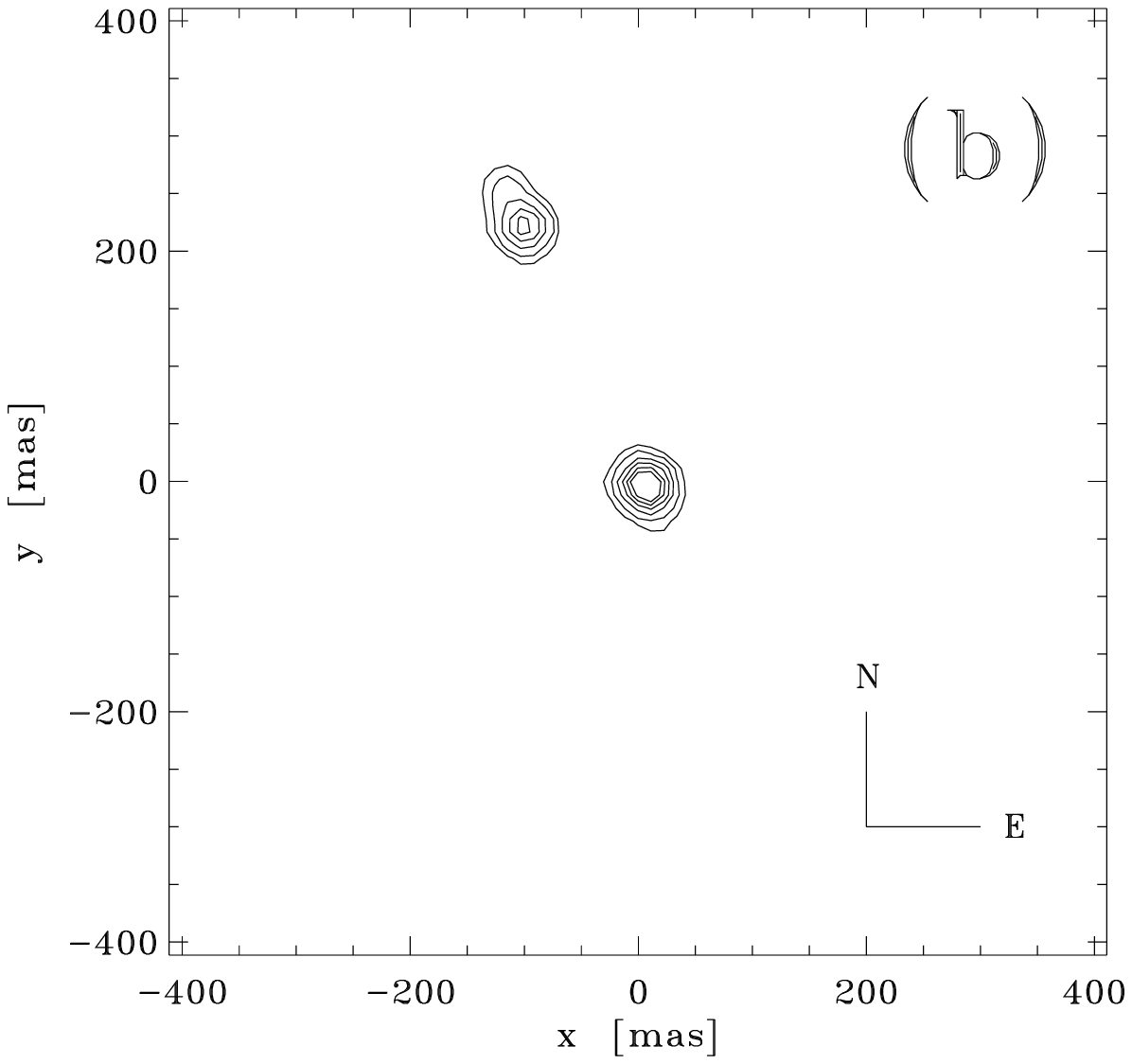}

\plottwo{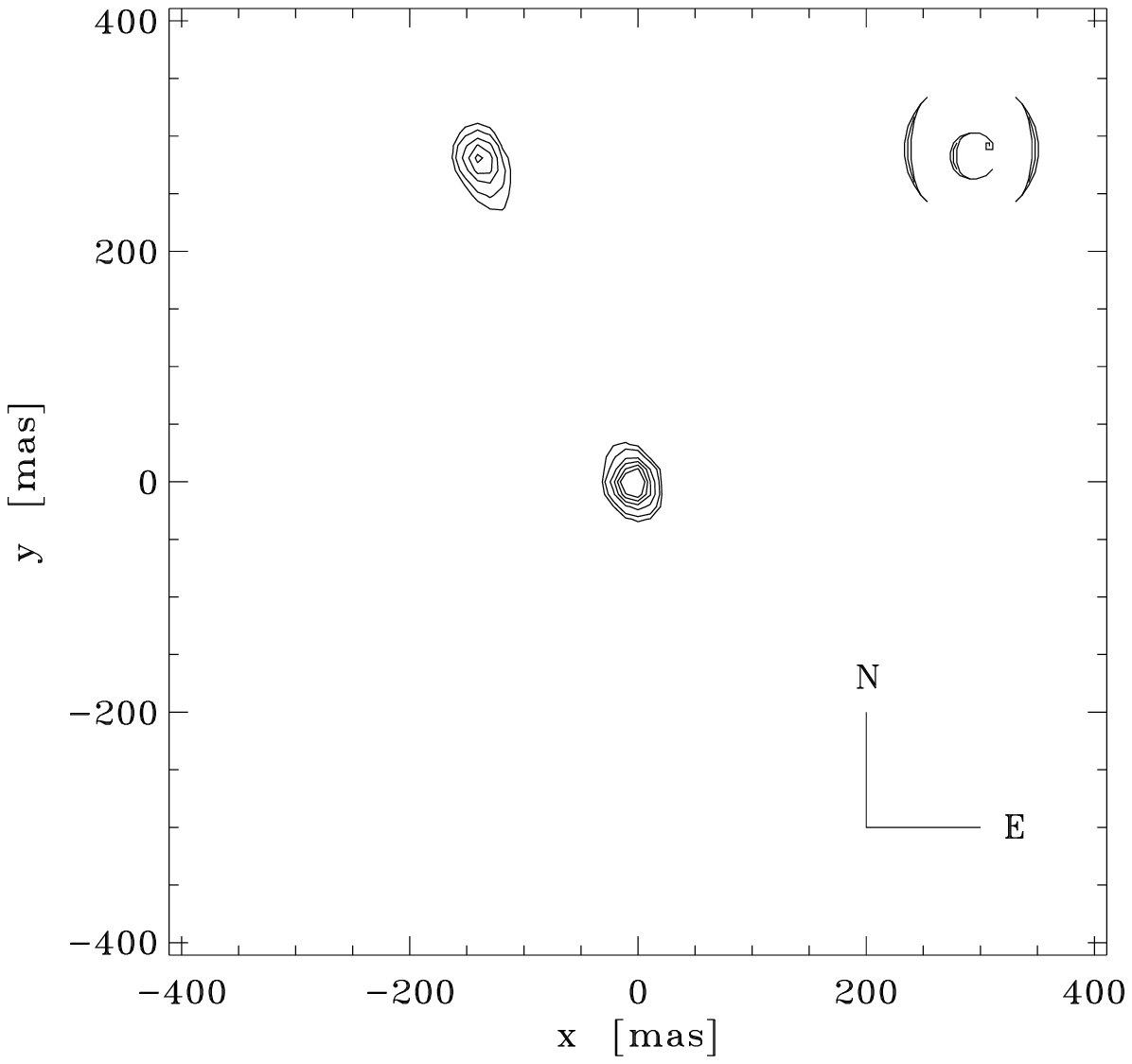}{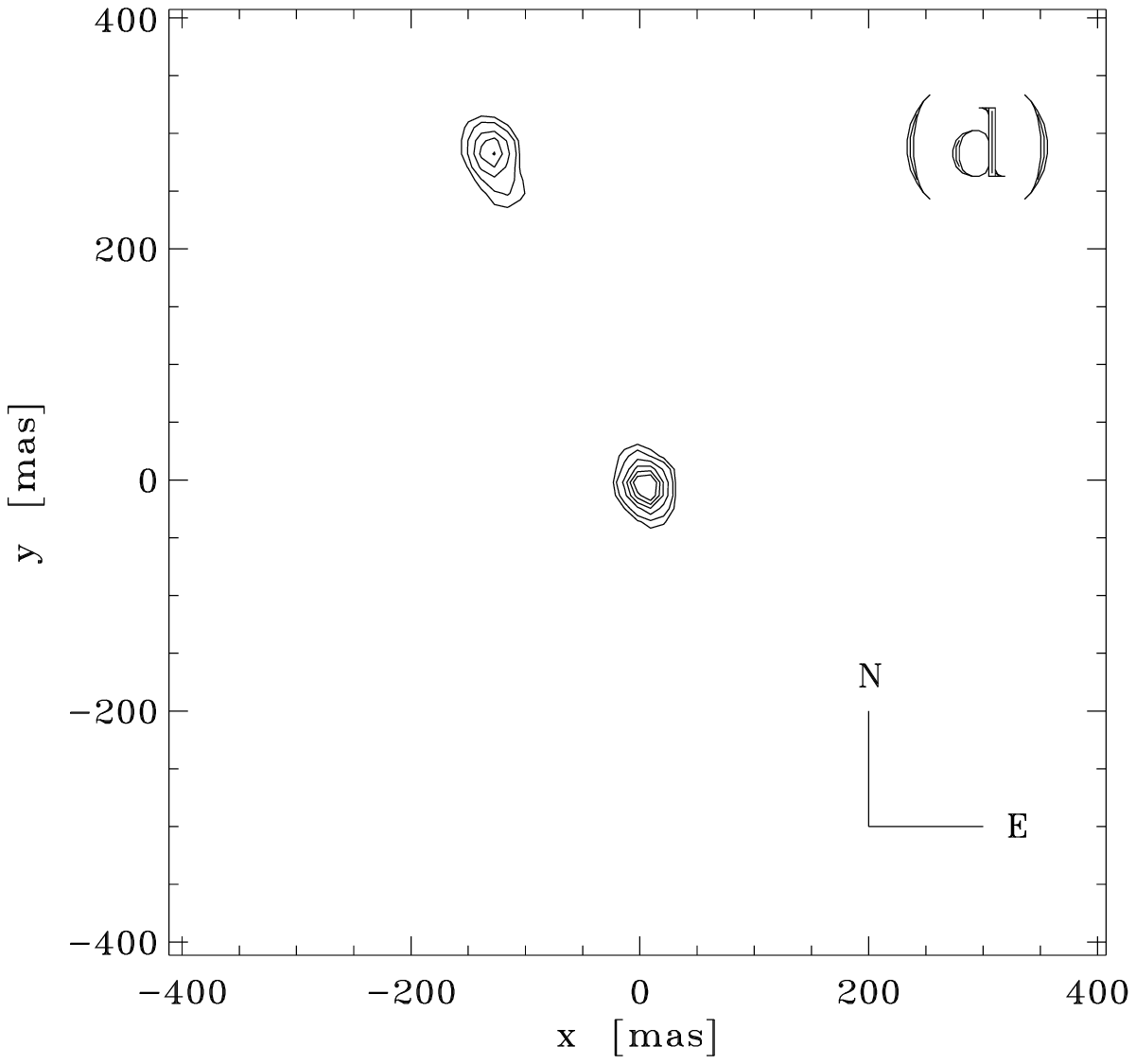}
\caption{
Contour plots of reconstructed images for HIP 111805, a triple system
where the secondary is a known spectroscopic binary with period 551.6 days.
Contours are drawn at 0.05, 0.1, 0.2, 0.3, 0.4, and 0.5 of the maximum
of each array (the central peak corresponding to the primary star). The images
labeled (a) and (b) are the 692- and 880-nm images from 27 July 2013
respectively, and
the images labeled (c) and (d) are the same for the observation of 25 July 
2014. The strong asymmetry shape of the contours of the secondary star
indicates that it is itself binary, 
and that the position angle has changed by approximately
180$^{\circ}$ from 2013 to 2014.
}
\end{figure}

\begin{figure}[tb]
\plottwo{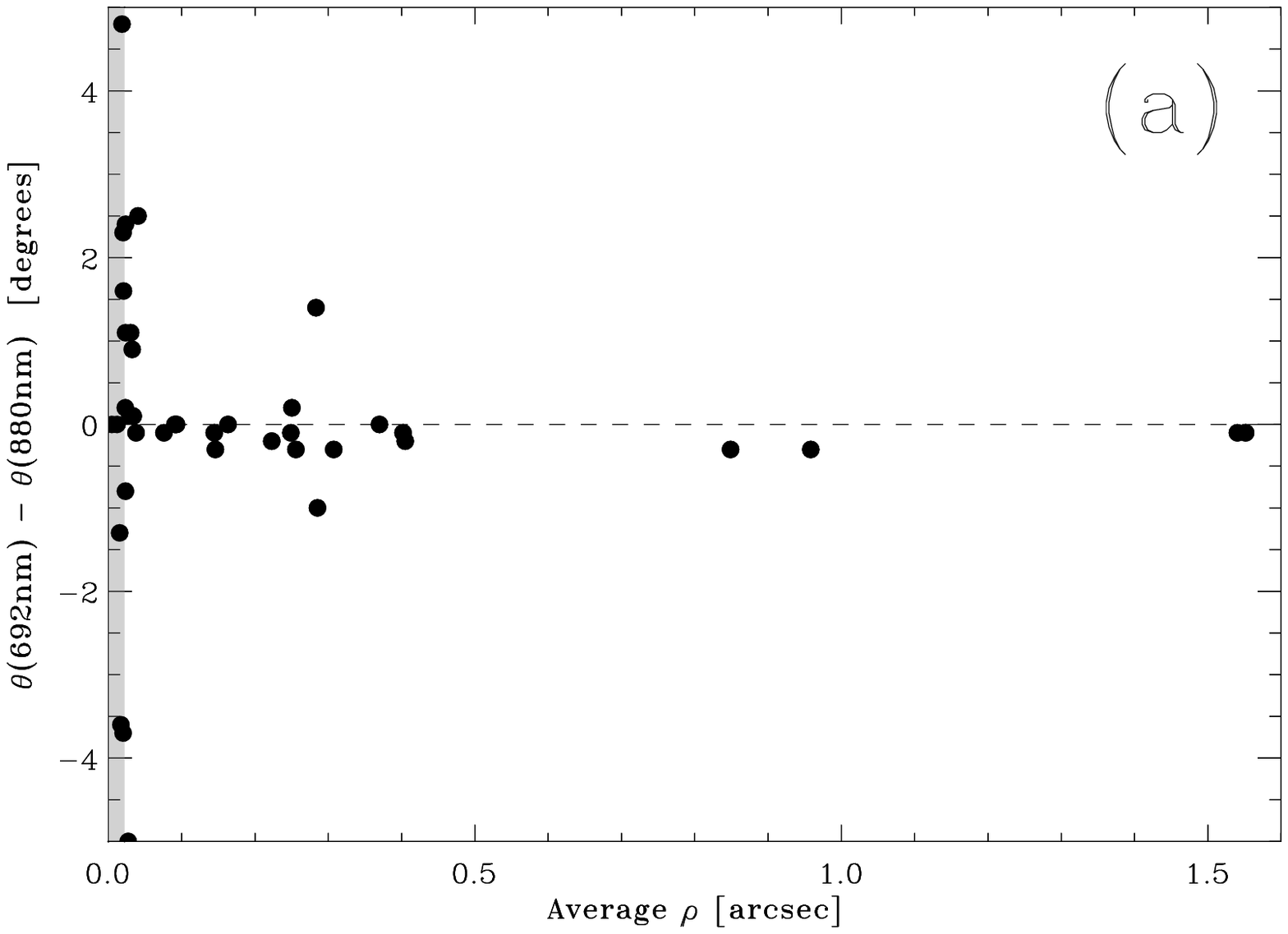}{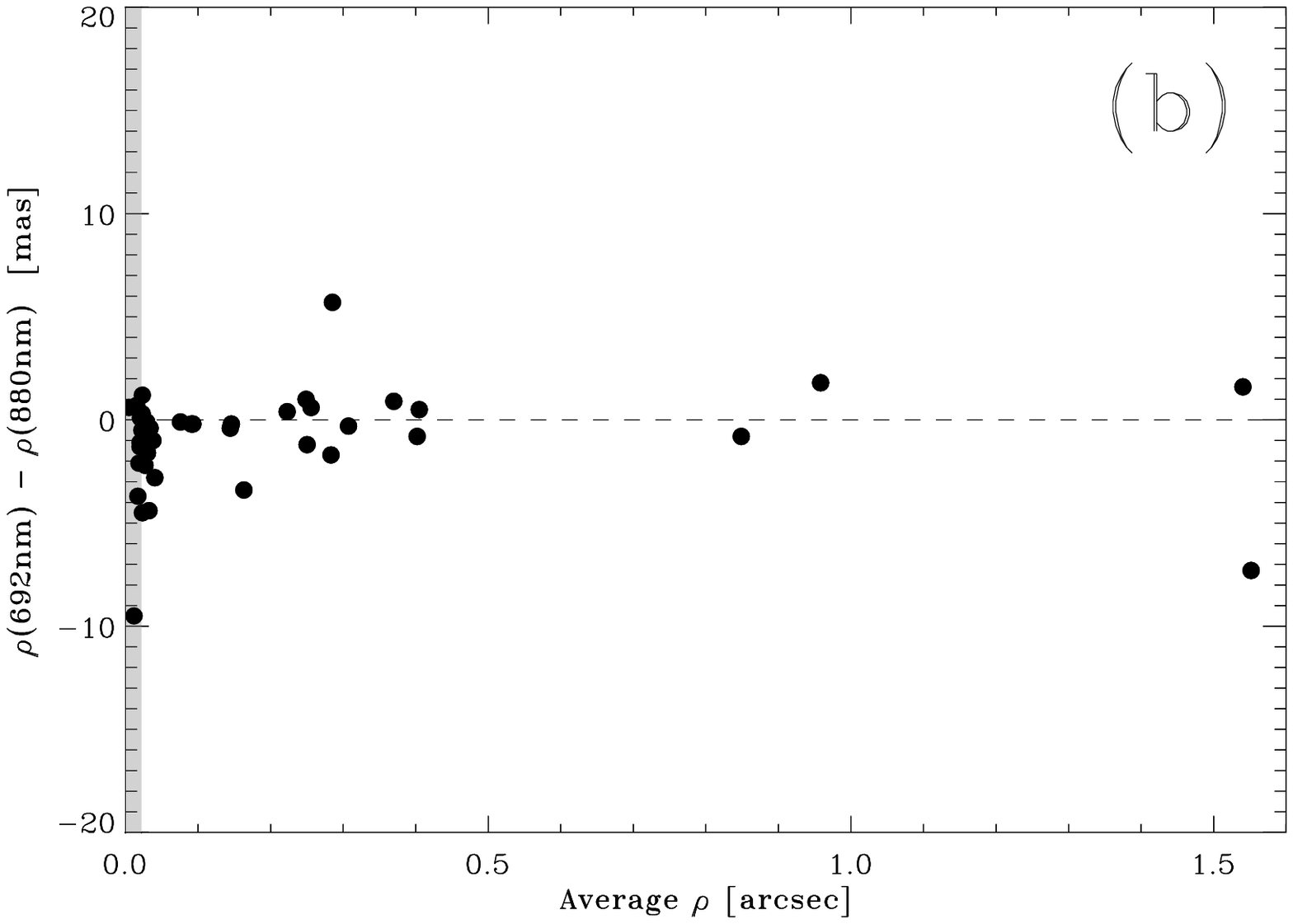}
\caption{
Measurement differences between the two channels of the instrument
plotted as a function of measured separation, $\rho$.
(a) Position angle ($\theta$) differences.
(b) Separation ($\rho$) differences.
In both plots, the gray band at the left marks the region below the
diffraction limit of the telescope.
}
\end{figure}

\begin{figure}[tb]
\plottwo{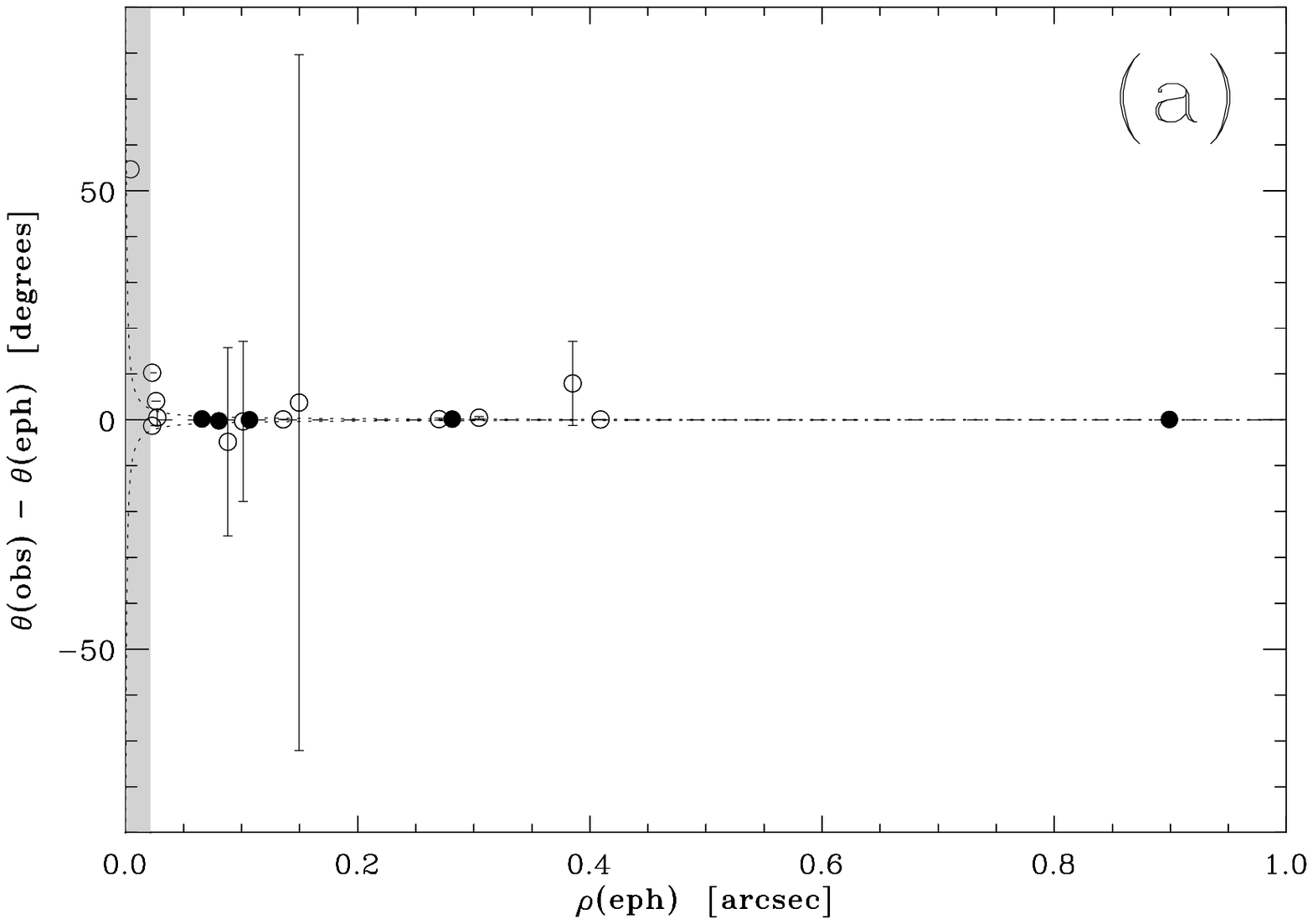}{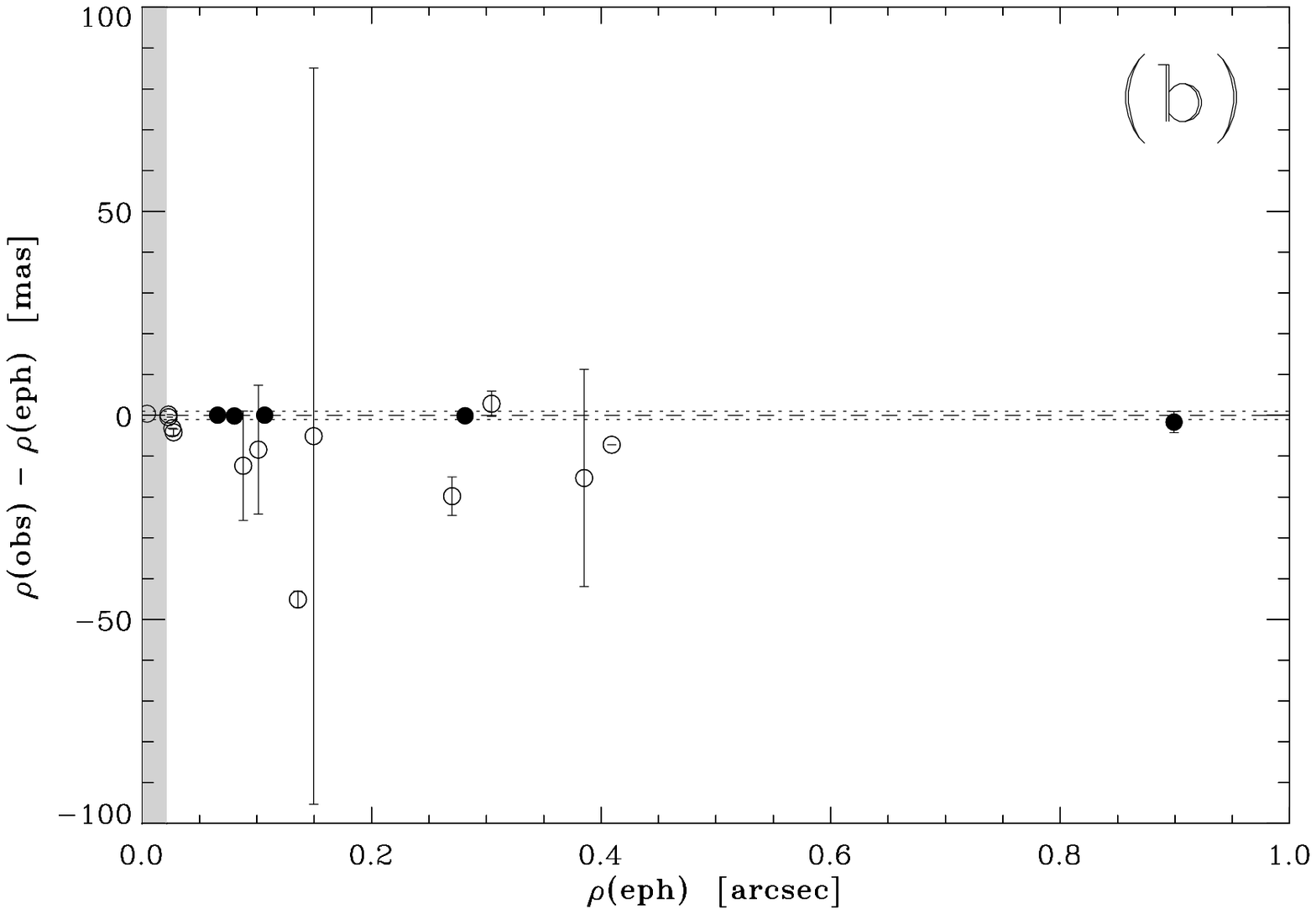}
\caption{
Observed minus ephemeris differences in position angle and separation
when comparing the measures presented here with orbital ephemerides
of objects having an orbit in the Sixth
Orbit Catalog of Hartkopf, Mason and Worley (2001). Observations in 
both channels have been averaged before subtracting the ephemeris value.
(a) Position angle residuals. In this plot, the dotted curves mark the
position angle error expected from a linear measurement error
of 1 mas, the derived value when averaging the values obtained from both
channels of the instrument as discussed in the text.
(b) Separation residuals.
The dotted line is drawn at 1 mas. 
In both plots, the gray region at the left of the plot marks the diffraction
limit of the 692 nm filter, and
the error bars indicate the uncertainties in the ephemeris position
based on error propagation of the published uncertainties in the
orbital elements. Objects with no published uncertainties are shown with
a horizontal line through the data point. Filled circles indicate
the objects used in the scale calibration.
}
\end{figure}

\begin{figure}[tb]
\plottwo{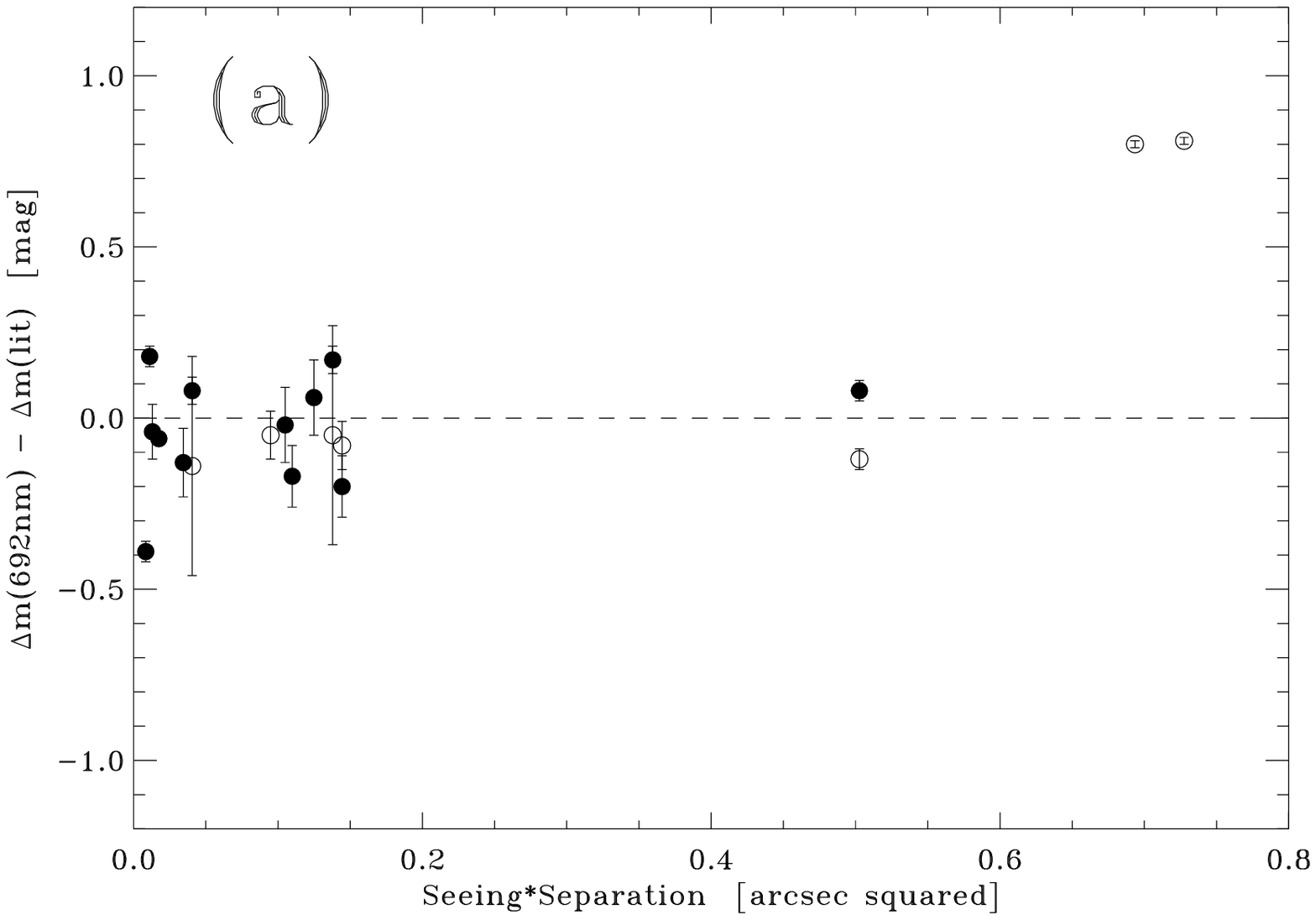}{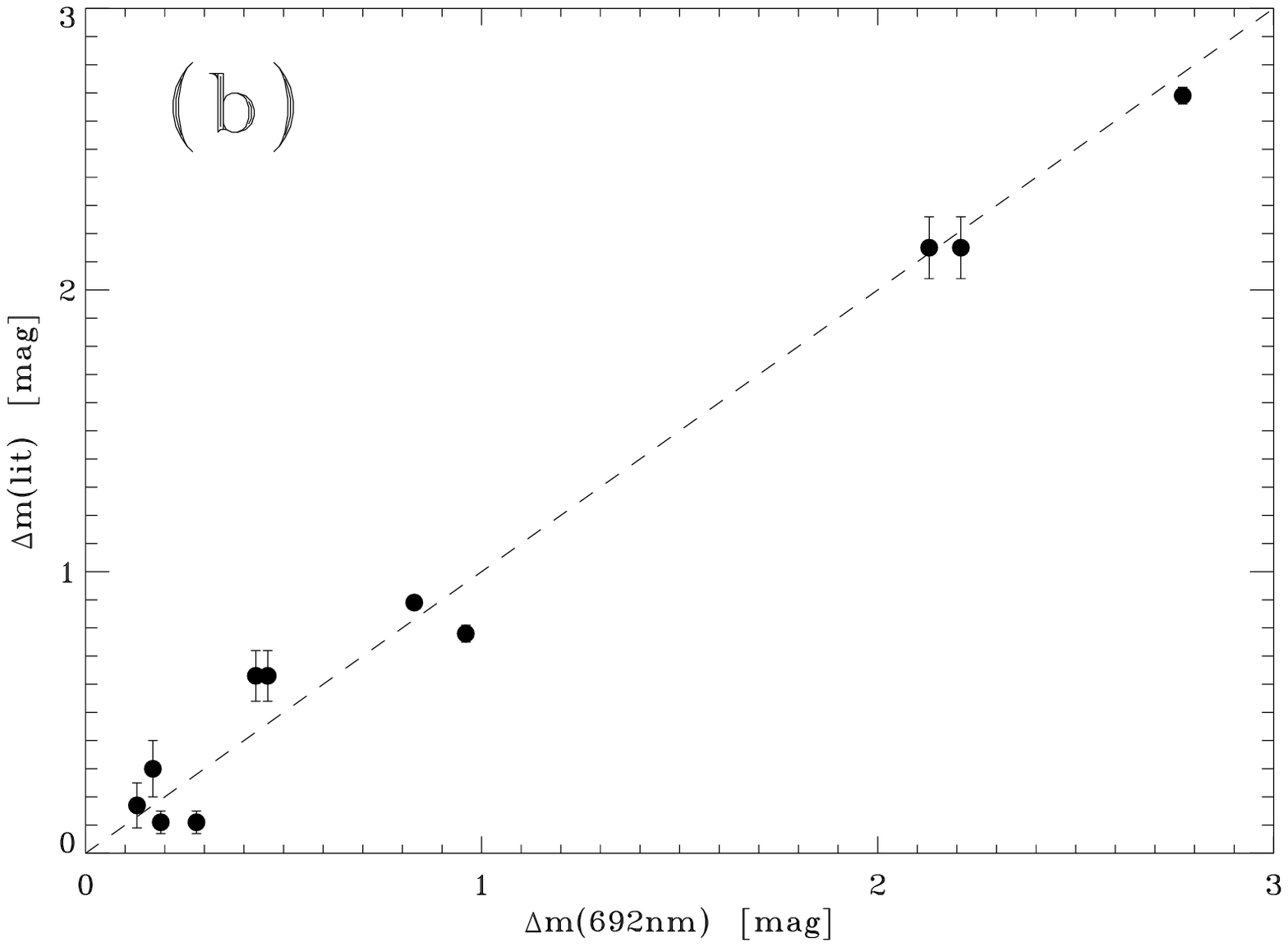}
\caption{
A comparison of the differential photometry presented in Table 1 with
existing differential photometry.
(a) The difference in $\Delta m$ between our measure at 692 nm and the
$\Delta m$ value appearing in the {\it Hipparcos} Catalogue as
a function of the parameter $q^{\prime}$ = seeing times
separation discussed in the text. 
Filled circles indicate the average of previous measures 
in the 4th Interferometric Catalog observed with a similar
filter to 692-nm, using the standard error in the mean as the error bar.
Open circles show the difference between our 692-nm result and that appearing
in the {\it Hipparcos} Catalogue, in the $H_{p}$ filter, with the error
bar being the uncertainty shown in the Catalogue.
To minimize color effects arising from the difference in filter wavelength
between our observations and {\it Hipparcos}, only systems with $B-V < 0.7$
are included for the Hipparcos comparison.
(b) A plot of the $\Delta m$ value as a function of the magnitude
difference at 692 nm in Table 1 for those systems with data in a 
similar filter in the 4th Interferometric Catalog and $q^{\prime} < 0.6$.
}
\end{figure}

\begin{figure}[tb]
\plottwo{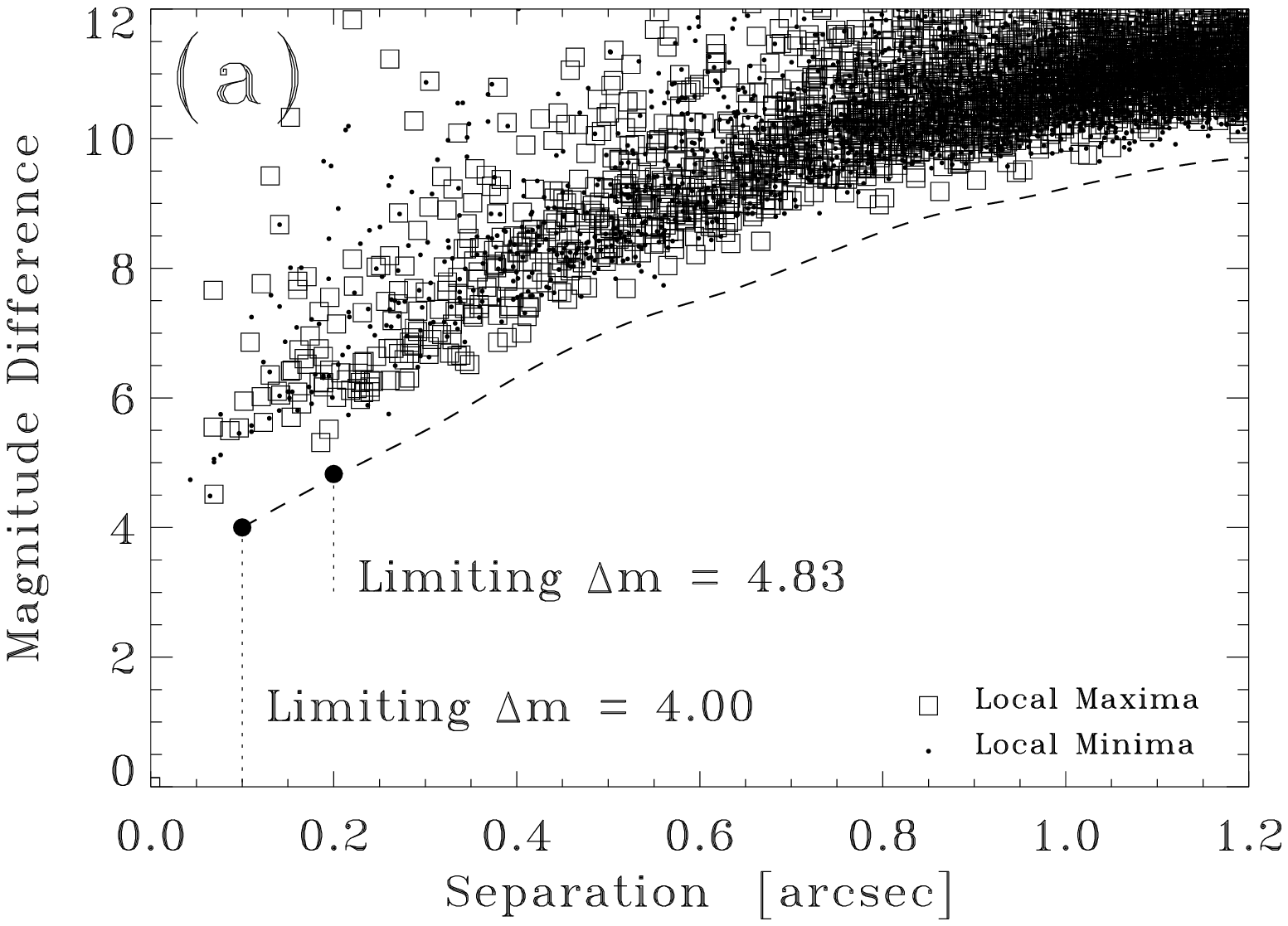}{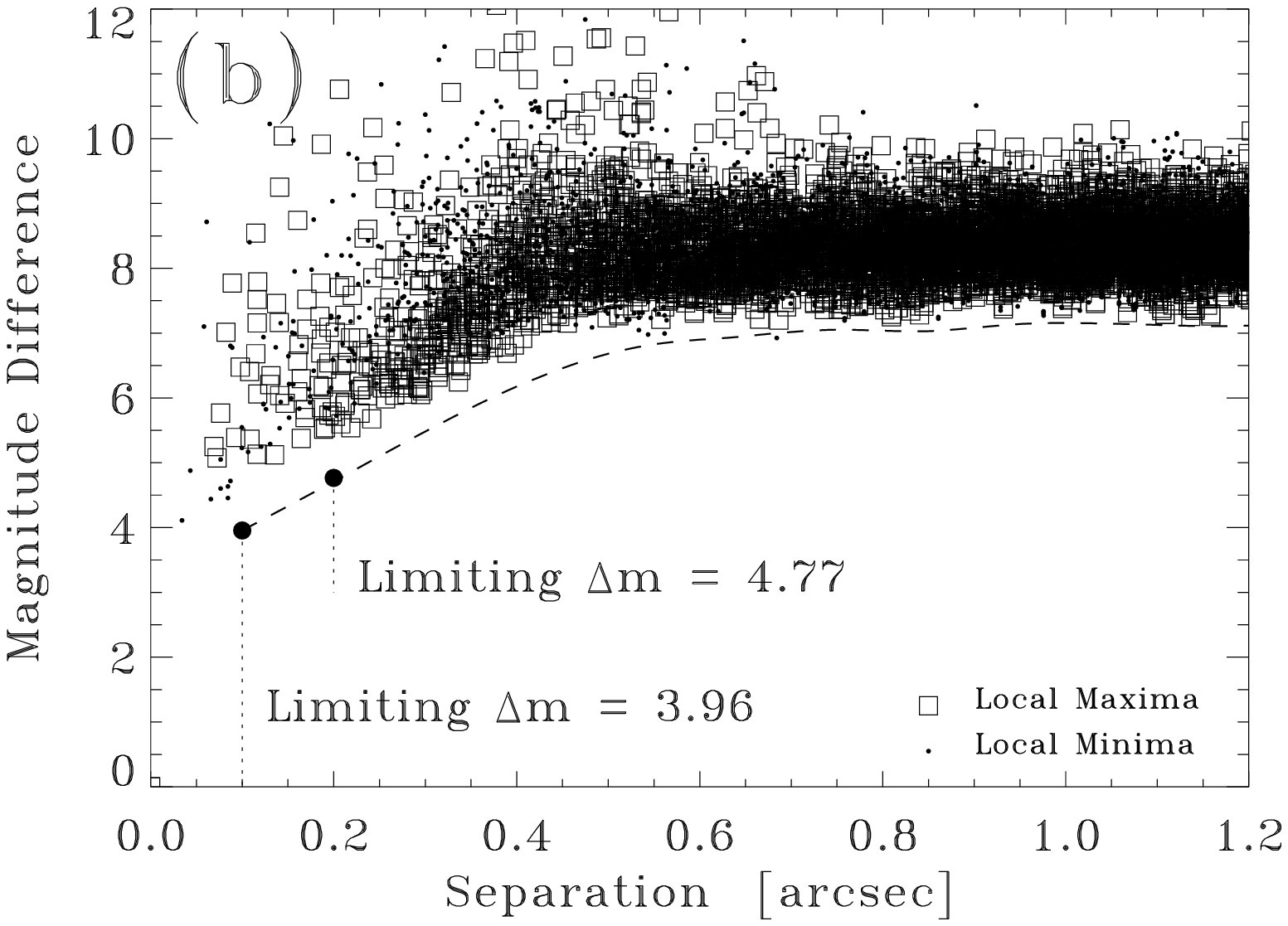}
\caption{
Detection limit curves for HIP 6917, a single-lined spectroscopic
binary star with a 10-day period. No secondary was detected in this case, to
the limit shown in each case.
(a) The result for the 692 nm reconstructed image.
(b) The result at 880 nm.
}
\end{figure}

\clearpage

\begin{figure}[tb]
\plottwo{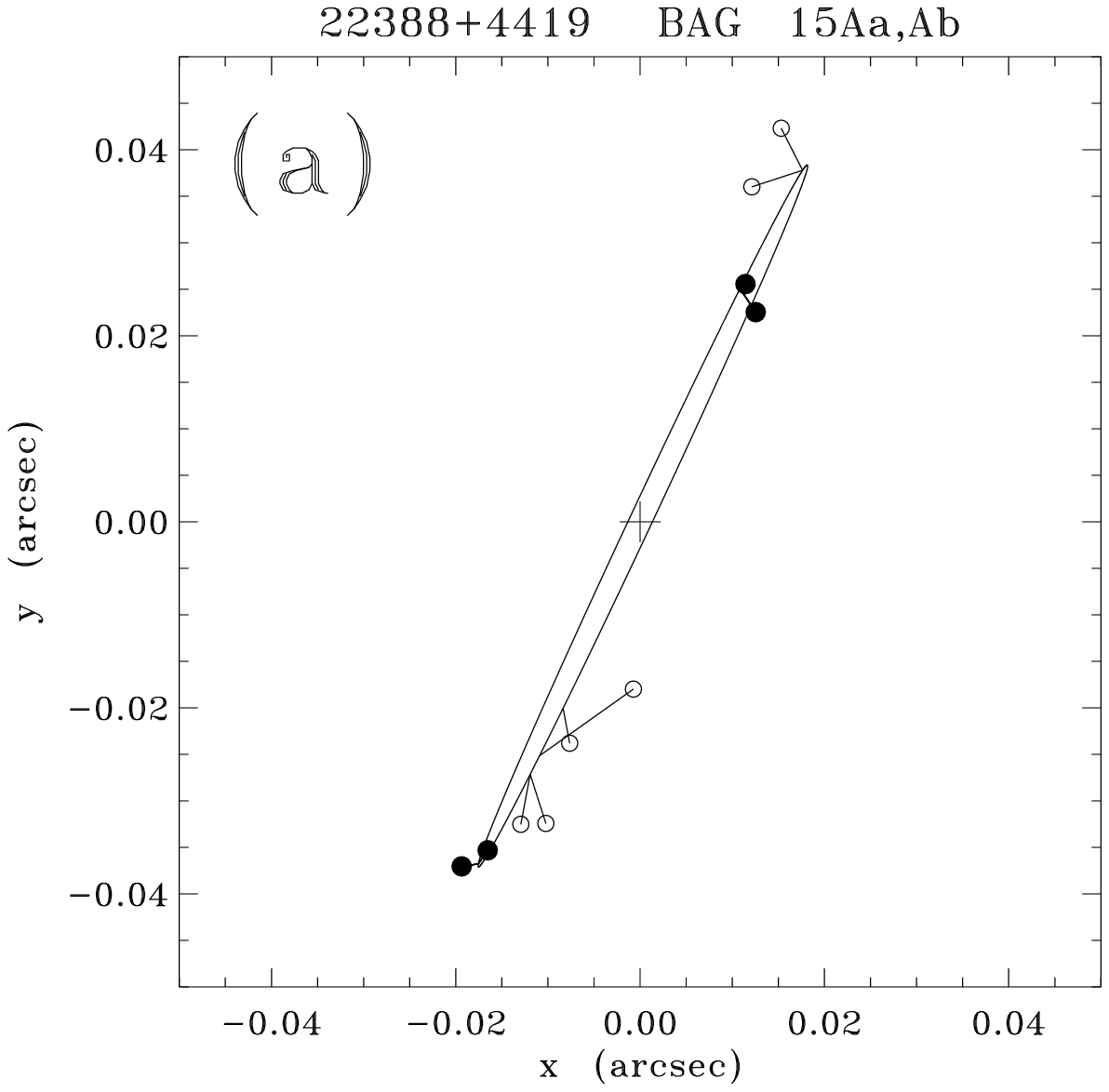}{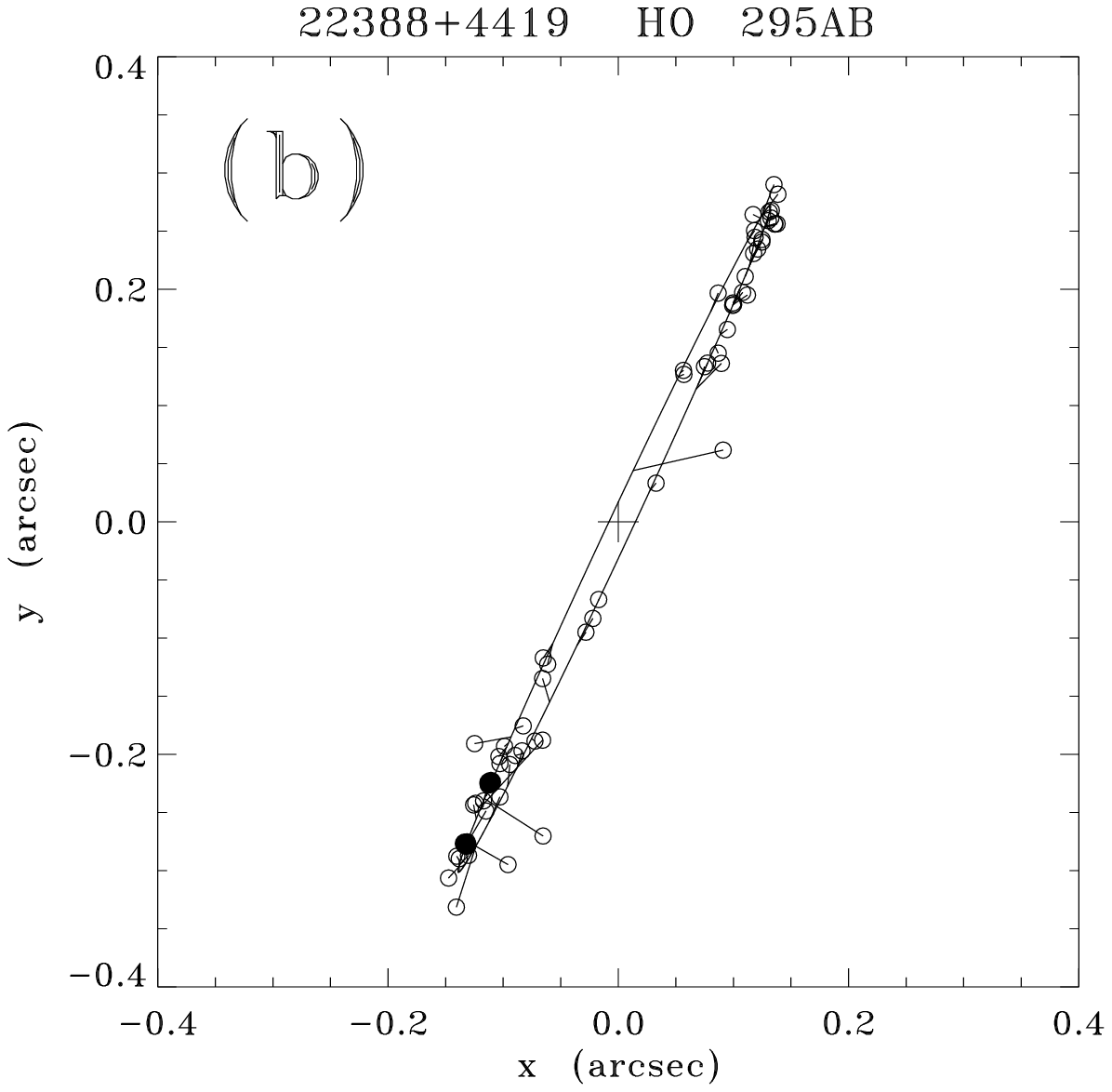}

\plottwo{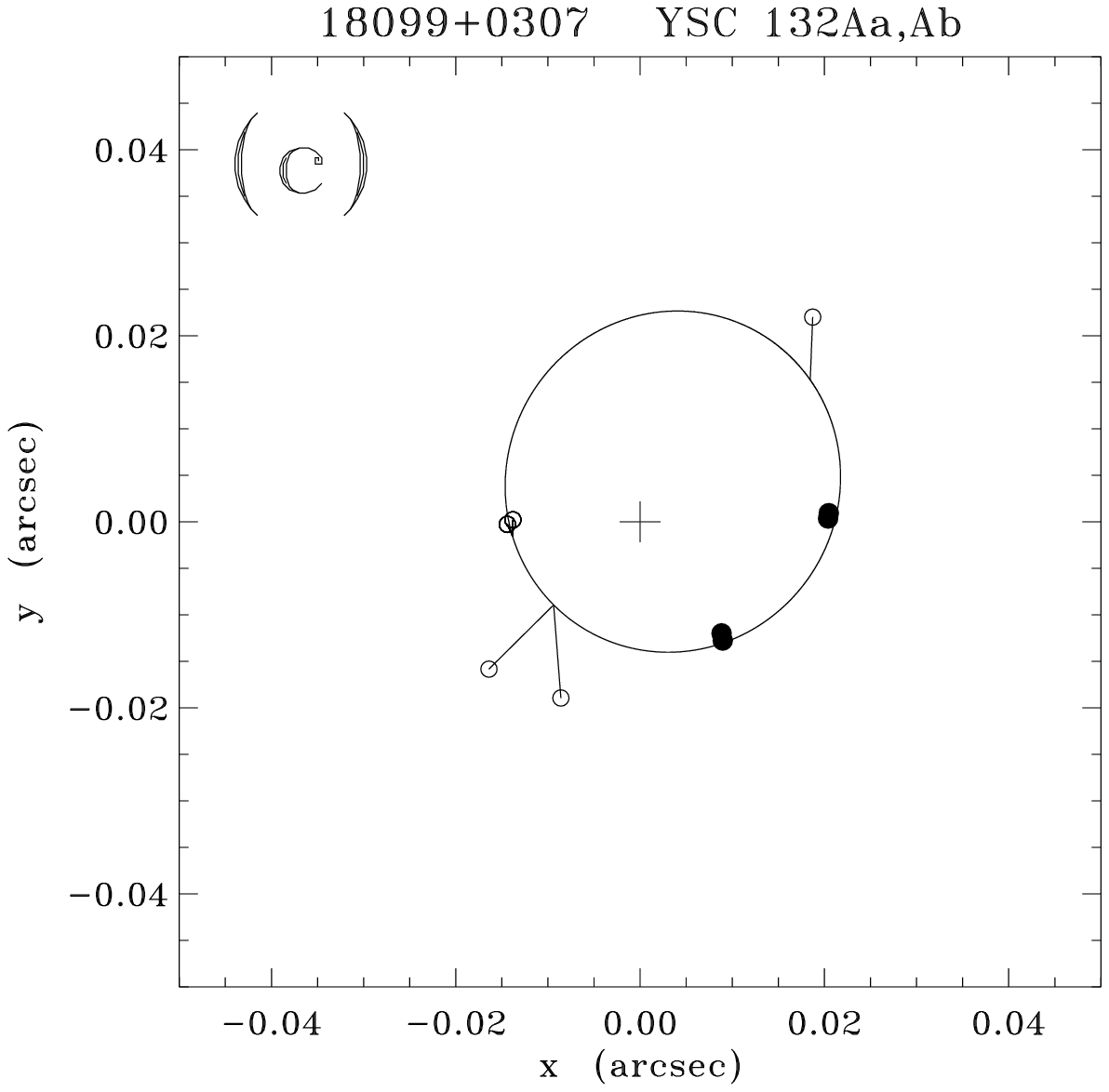}{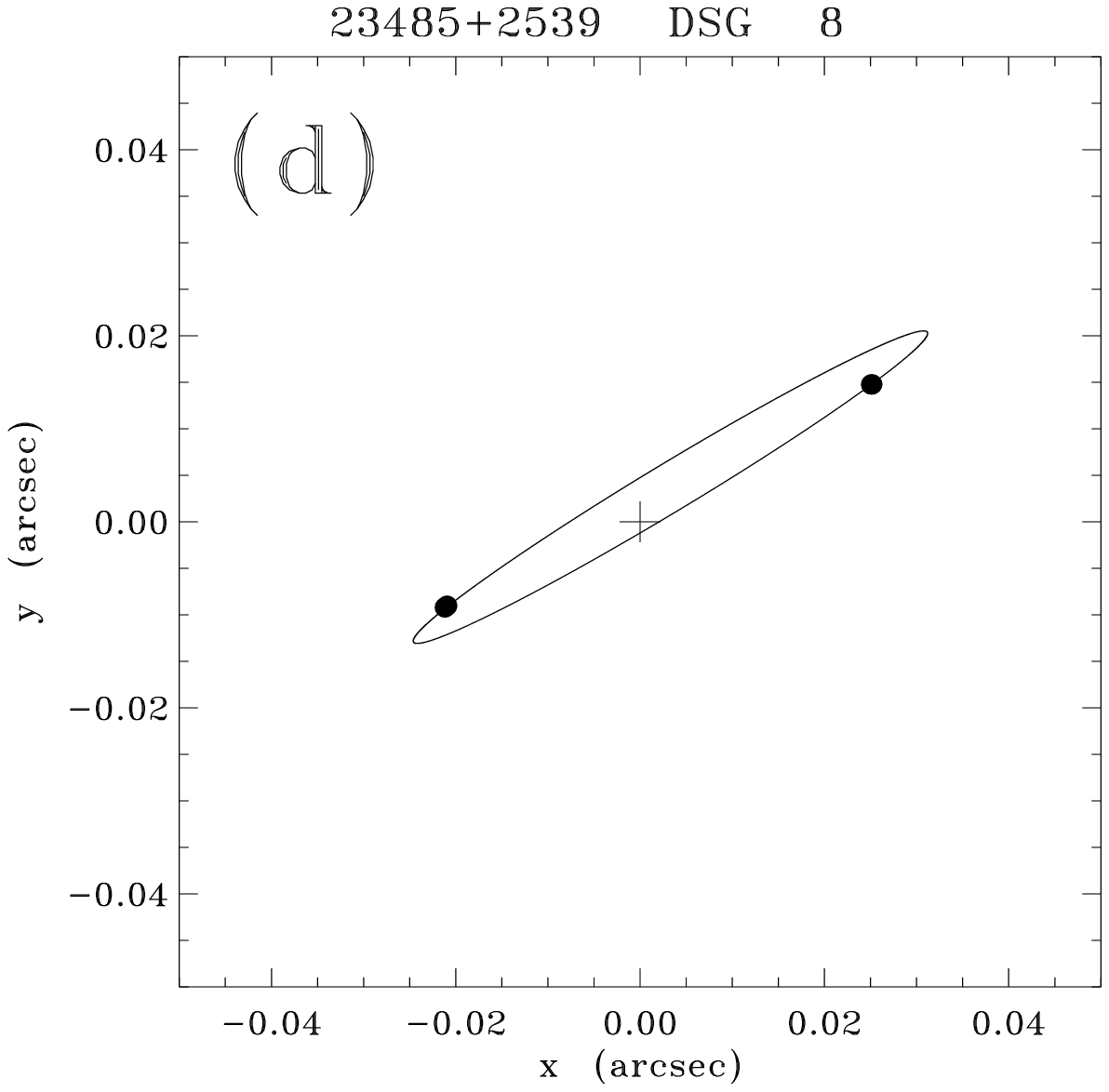}
\caption{
The orbits calculated here for four systems
together with data from the literature
and our measures from Table 1. The latter are shown with filled circles.
All points are drawn with line segments from the data point to the
location of the ephemeris prediction on the orbital path.
(a) BAG 15Aa,Ab (= HIP 111805),
(b) HO 295AB (= HIP 111805); note that (a) and (b) are a triple system.
(c) YSC 132Aa,Ab (= HIP 89000); (d) DSG 8 (= HIP 117415).
North is down and east is to the right in all cases.
}
\end{figure}

\begin{figure}[tb]
\plotone{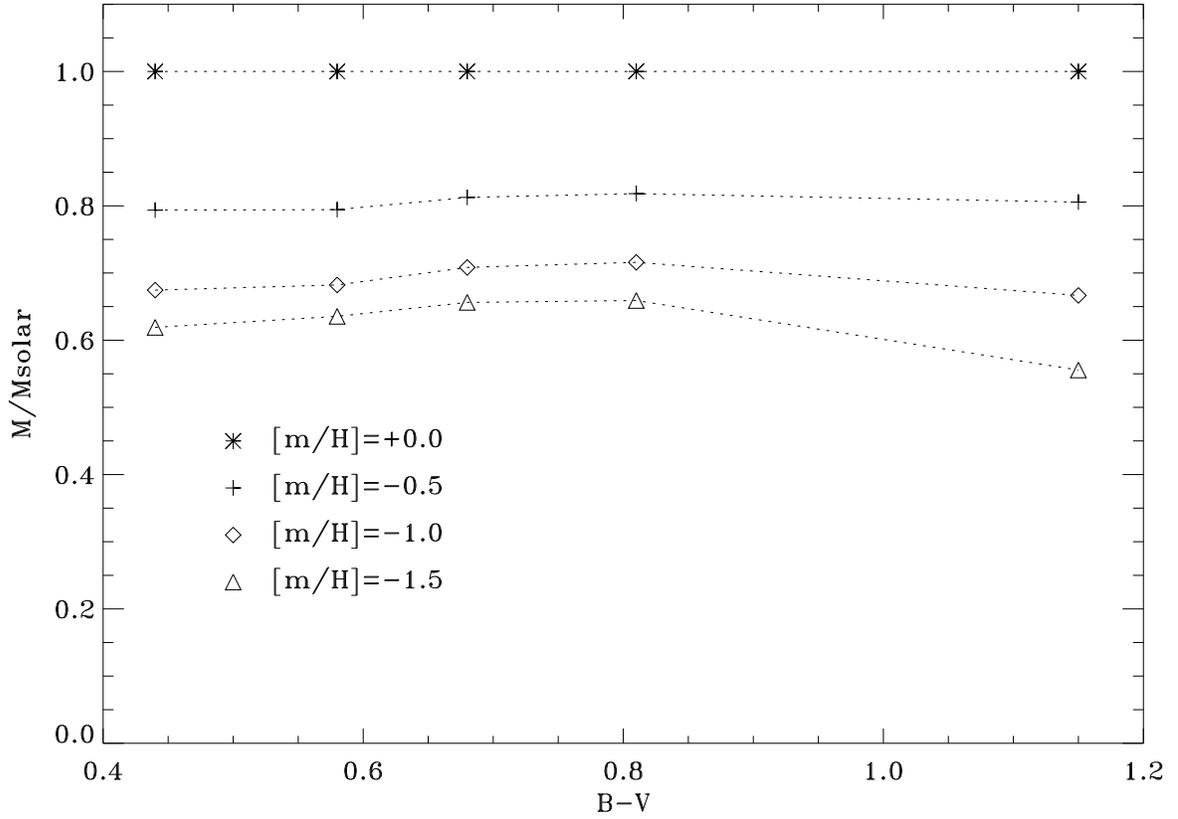}
\caption{
The behavior of stellar mass relative to the solar metallicity
value as a function
of $B-V$ and metal abundance, as predicted by the Spada \ea (2013)
isochrones. 
This example is for a mixing length parameter of 1.875 and an age of 0.1 Gyr,
but other ages and mixing lengths give similar results.
For a given metal abundance, the mass ratio is relatively uniform across
the spectral range of interest.
}
\end{figure}

\begin{figure}[tb]
\plottwo{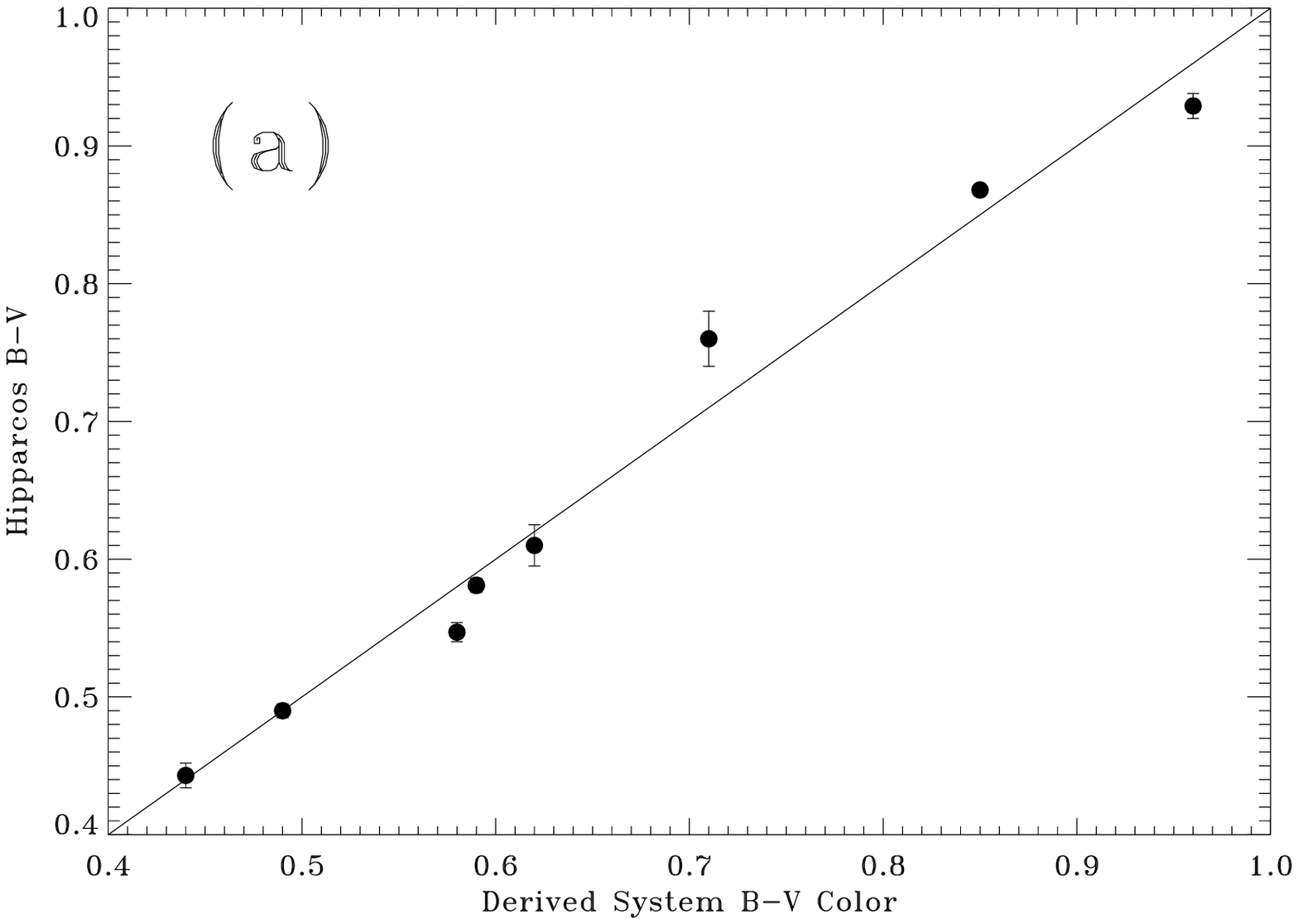}{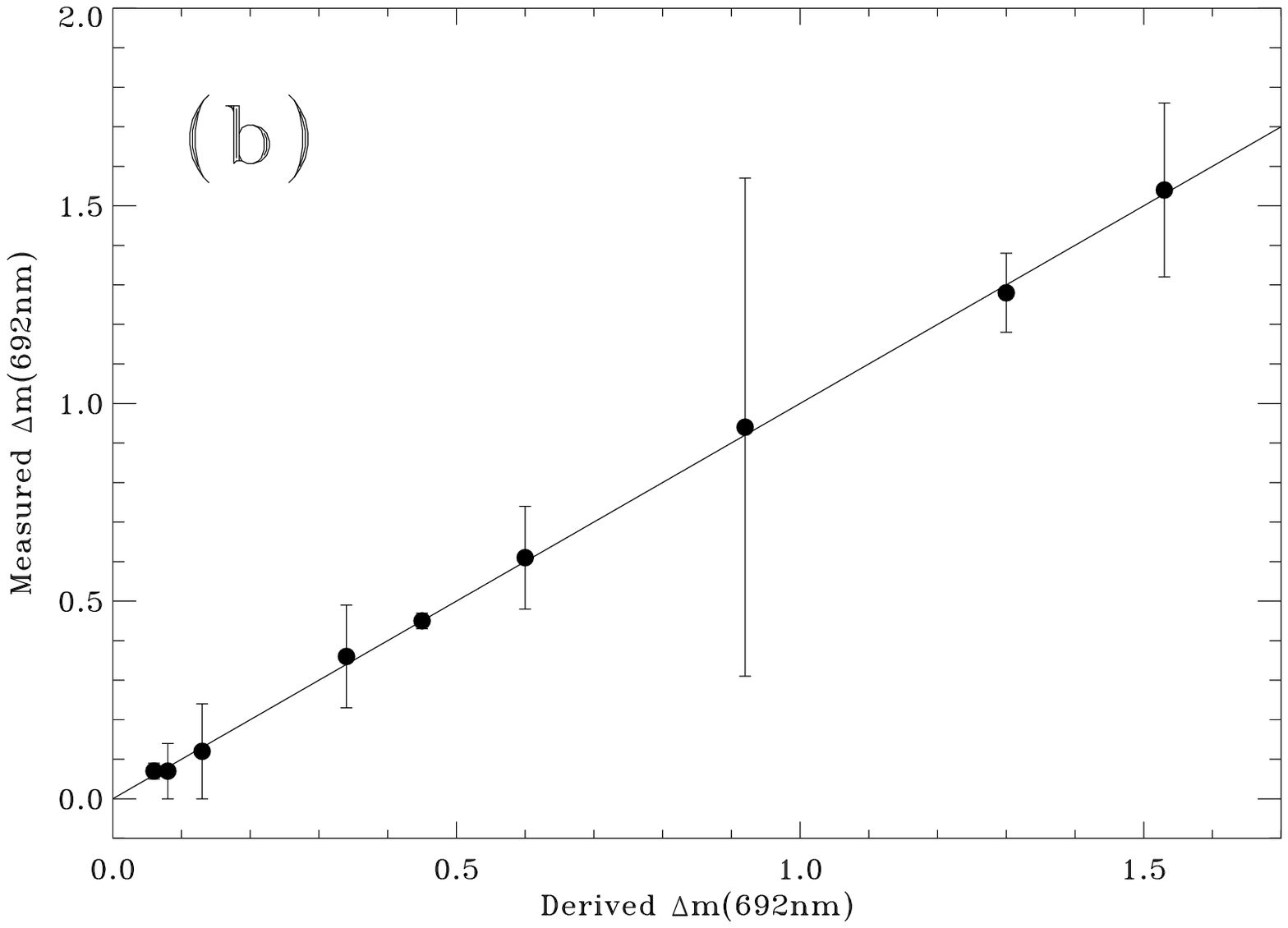}
\caption{
Measured properties of the systems versus the properties derived
from the Pickles-based simulations.
(a) $B-V$ color.
(b) Magnitude difference at 692 nm.
In both cases, the line drawn is $y=x$, indicating that the simulated
results are in good agreement with the observed quantities.
}
\end{figure}

\begin{figure}[tb]
\plotone{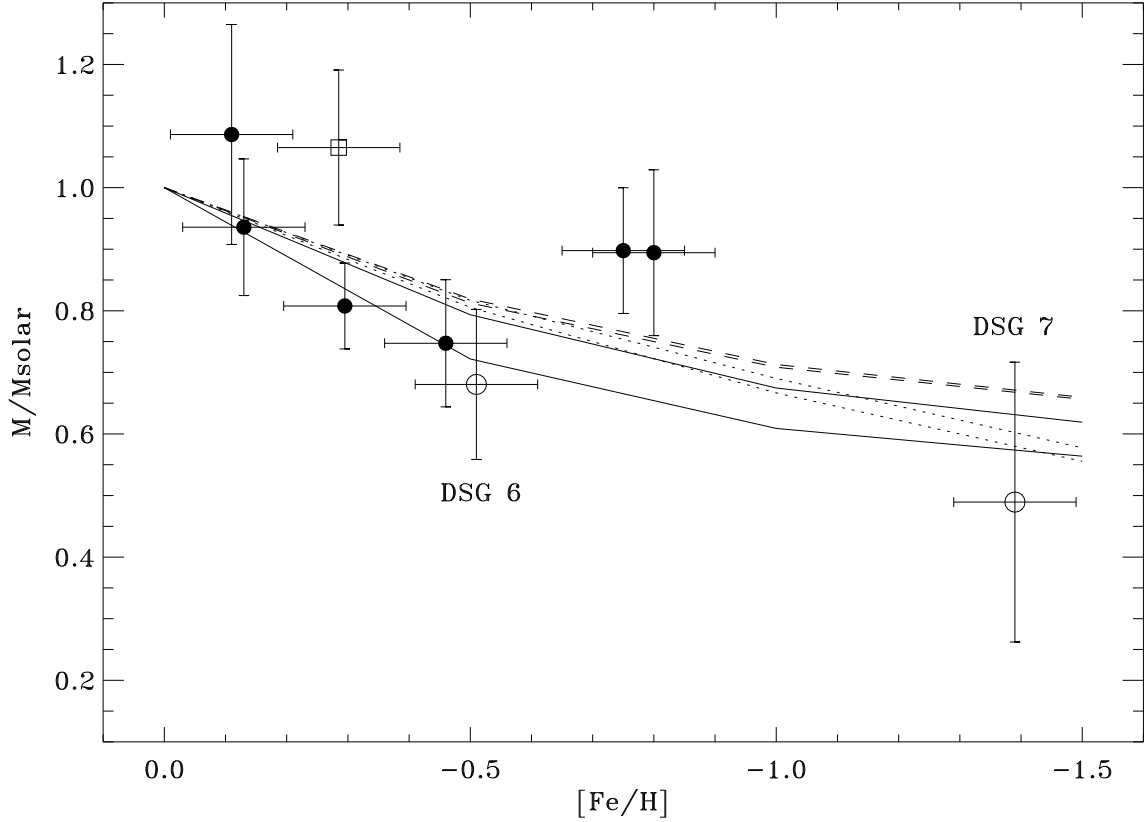}
\caption{
Simulated solar metallicity binaries are constructed with components that match the $B-V$ 
colors and observed magnitude differences for a set of nine sub-solar metallicity 
binaries. The ratio of the total mass of each observed binary to the total mass of its
(solar-metallicity) simulated counterpart is plotted as a function of observed
metallicity.
The curves shown are derived from the Spada \ea (2013) stellar
models as described in the text; 
solid lines correspond to a spectral type of F5V ($B-V = 0.44$),
dashed curves to G5V ($B-V = 0.68$), and dotted lines to K5V ($B-V = 1.15$). In
all three cases, curves for two ages are shown, 0.1 Gyr and 4.0 Gyr.
The single-lined spectroscopic binaries
in Table 3 and 4 are shown as open circles, and the double-lined systems
are shown as filled circles. BAG 15Aa,Ab is shown as with an open square;
this system is discussed in the text as having a discrepancy
between the time of periastron passage as determined from spectroscopy 
versus the result from the relative astrometry. 
The observed trend toward smaller
masses as the metallicity 
decreases matches the prediction from the stellar models.
}
\end{figure}

\end{document}